\documentclass[longauth]{aa}
\usepackage{graphicx}
\usepackage{txfonts}
\usepackage{booktabs}

\usepackage{fontawesome}
\usepackage{xcolor}
\usepackage{lineno}

\usepackage[breaklinks, colorlinks, citecolor=blue, linkcolor=blue]{hyperref}
\usepackage[normalem]{ulem}

\newcommand{\ud}{\ensuremath{\mathrm{d}\xspace}}

\newcommand{\nstar}{KELT-1\xspace}
\newcommand{\nplanet}{KELT-1b\xspace}

\newcommand{\pytransit}{\textsc{PyTransit}\xspace}
\newcommand{\ldtk}{\textsc{LDTk}\xspace}
\newcommand{\george}{\textsc{George}\xspace}
\newcommand{\celerite}{\textsc{Celerite}\xspace}

\newcommand{\tess}{\textit{TESS}\xspace}
\newcommand{\spitzer}{\textit{Spitzer}\xspace}

\newcommand{\cheops}{\textit{CHEOPS}\xspace}

\newcommand{\teff}{\ensuremath{T_\mathrm{eff}}\xspace}
\newcommand{\tbr}{\ensuremath{T_\mathrm{b}}\xspace}
\newcommand{\gcm}{\ensuremath{\mathrm{g\,cm^{-3}}}}

\newcommand{\rjup}{\ensuremath{R_\mathrm{Jup}}\xspace}
\newcommand{\msun}{\ensuremath{M_\odot}\xspace}
\newcommand{\rsun}{\ensuremath{R_\odot}\xspace}

\newcommand{\ag}{\ensuremath{A_\mathrm{g}}\xspace}
\newcommand{\agb}{\ensuremath{A_\mathrm{g,0}}\xspace}
\newcommand{\agr}{\ensuremath{A_\mathrm{g,1}}\xspace}

\newcommand{\pbh}{\textit{H}\xspace}
\newcommand{\pbks}{\textit{Ks}\xspace}
\newcommand{\pbsa}{3.6~$\mu$m\xspace}
\newcommand{\pbsb}{4.5~$\mu$m\xspace}

\makeatletter
\renewcommand*\aa@pageof{, page \thepage{} of \pageref*{LastPage}}
\def\instrefs#1{{\def\scsep{\def\scsep{,}}\@for\w:=#1\do{\scsep\ref{inst:\w}}}}
\makeatother

\newcommand{\ghlink}[2]{\href{https://github.com/hpparvi/cheops_kelt_1/tree/master/#1}{#2~\faicon{github}}}
\newcommand{\lpfed}{\ghlink{src/externaldatalpf.py}{\texttt{ExternalDataLPF}}\xspace}
\newcommand{\lpffn}{\ghlink{src/finallpf.py}{\texttt{FinalLPF}}\xspace}

\newcommand{\orcidicon}[1]{
  \href{https://orcid.org/#1}{\includegraphics[scale=0.16]{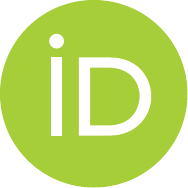}}
}

\begin{document} 
  \title{\cheops finds \nplanet darker than expected in visible light}
  \subtitle{Discrepancy between the \cheops and \tess eclipse depths}
  \author{
H. Parviainen\inst{\ref{inst:1},\ref{inst:ull}}\orcidicon{0000-0001-5519-1391} \and
T. G. Wilson\inst{\ref{inst:2}}\orcidicon{0000-0001-8749-1962} \and
M. Lendl\inst{\ref{inst:3}}\orcidicon{0000-0001-9699-1459} \and
D.~Kitzmann\inst{\ref{inst:5}}\orcidicon{0000-0003-4269-3311} \and
E.~Pallé\inst{\ref{inst:1},\ref{inst:ull}}\orcidicon{0000-0003-0987-1593} \and
L. M. Serrano\inst{\ref{inst:10}}\orcidicon{0000-0001-9211-3691} \and
E.~Meier Valdes\inst{\ref{inst:5}}\orcidicon{0000-0002-2160-8782 } \and 
W. Benz\inst{\ref{inst:4},\ref{inst:5}}\orcidicon{0000-0001-7896-6479} \and
A. Deline\inst{\ref{inst:3}} \and 
D.~Ehrenreich\inst{\ref{inst:3}}\orcidicon{0000-0001-9704-5405} \and
P. Guterman\inst{\ref{inst:6},\ref{inst:7}} \and 
K. Heng\inst{\ref{inst:5},\ref{inst:13}}\orcidicon{0000-0003-1907-5910} \and
O.~D.~S.~Demangeon\inst{\ref{inst:8},\ref{inst:9}}\orcidicon{0000-0001-7918-0355} \and
A.~Bonfanti\inst{\ref{inst:11}}\orcidicon{0000-0002-1916-5935} \and
S.~Salmon\inst{\ref{inst:3}}\orcidicon{0000-0002-1714-3513} \and
V. Singh\inst{\ref{inst:12}}\orcidicon{0000-0002-7485-6309} \and
N. C. Santos\inst{\ref{inst:8},\ref{inst:9}}\orcidicon{0000-0003-4422-2919} \and
S.~G.~Sousa\inst{\ref{inst:8}}\orcidicon{0000-0001-9047-2965} \and
Y.~Alibert\inst{\ref{inst:4}}\orcidicon{0000-0002-4644-8818} \and
R.~Alonso\inst{\ref{inst:1},\ref{inst:ull}}\orcidicon{0000-0001-8462-8126} \and
G. Anglada\inst{\ref{inst:15},\ref{inst:16}}\orcidicon{0000-0002-3645-5977} \and
T. Bárczy\inst{\ref{inst:17}}\orcidicon{0000-0002-7822-4413} \and
D. Barrado y Navascues\inst{\ref{inst:18}}\orcidicon{0000-0002-5971-9242} \and
S.~C.~C.~Barros\inst{\ref{inst:8},\ref{inst:9}}\orcidicon{0000-0003-2434-3625} \and
W.~Baumjohann\inst{\ref{inst:11}}\orcidicon{0000-0001-6271-0110} \and
M. Beck\inst{\ref{inst:3}}\orcidicon{0000-0003-3926-0275} \and
T. Beck\inst{\ref{inst:4}} \and 
N. Billot\inst{\ref{inst:3}}\orcidicon{0000-0003-3429-3836} \and
X.~Bonfils\inst{\ref{inst:19}}\orcidicon{0000-0001-9003-8894} \and
A.~Brandeker\inst{\ref{inst:20}}\orcidicon{0000-0002-7201-7536} \and
C. Broeg\inst{\ref{inst:4},\ref{inst:5}}\orcidicon{0000-0001-5132-2614} \and
J.~Cabrera\inst{\ref{inst:21}} \and 
S. Charnoz\inst{\ref{inst:22}}\orcidicon{0000-0002-7442-491X} \and
A. Collier Cameron\inst{\ref{inst:2}}\orcidicon{0000-0002-8863-7828} \and
C.~Corral Van Damme\inst{\ref{inst:29}} \and
Sz. Csizmadia\inst{\ref{inst:21}}\orcidicon{0000-0001-6803-9698} \and
M.~B.~Davies\inst{\ref{inst:23}}\orcidicon{0000-0001-6080-1190} \and
M. Deleuil\inst{\ref{inst:6}}\orcidicon{0000-0001-6036-0225} \and
L. Delrez\inst{\ref{inst:24},\ref{inst:25}}\orcidicon{0000-0001-6108-4808} \and
B.-O.~Demory\inst{\ref{inst:5}}\orcidicon{0000-0002-9355-5165} \and
A.~Erikson\inst{\ref{inst:21}} \and 
J.~Farinato\inst{\ref{inst:34}} \and
A.~Fortier\inst{\ref{inst:4},\ref{inst:5}}\orcidicon{0000-0001-8450-3374} \and
L.~Fossati\inst{\ref{inst:11}}\orcidicon{0000-0003-4426-9530} \and
M. Fridlund\inst{\ref{inst:26},\ref{inst:27}}\orcidicon{0000-0002-0855-8426} \and
D.~Gandolfi\inst{\ref{inst:10}}\orcidicon{0000-0001-8627-9628} \and
M.~Gillon\inst{\ref{inst:24}}\orcidicon{0000-0003-1462-7739} \and
M.~Güdel\inst{\ref{inst:28}} \and 
S. Hoyer\inst{\ref{inst:6}}\orcidicon{0000-0003-3477-2466} \and
K. G. Isaak\inst{\ref{inst:29}}\orcidicon{0000-0001-8585-1717} \and
L.~L.~Kiss\inst{\ref{inst:30},\ref{inst:31}} \and 
E.~Kopp\inst{\ref{inst:37}} \and
J.~Laskar\inst{\ref{inst:32}}\orcidicon{0000-0003-2634-789X} \and
A.~Lecavelier des Etangs\inst{\ref{inst:33}}\orcidicon{0000-0002-5637-5253} \and
C.~Lovis\inst{\ref{inst:3}}\orcidicon{0000-0001-7120-5837} \and
D.~Magrin\inst{\ref{inst:34}}\orcidicon{0000-0003-0312-313X} \and
P.~F.~L.~Maxted\inst{\ref{inst:35}}\orcidicon{0000-0003-3794-1317} \and
M.~Mecina\inst{\ref{inst:36}} \and
V.~Nascimbeni\inst{\ref{inst:34}}\orcidicon{0000-0001-9770-1214} \and
G.~Olofsson\inst{\ref{inst:20}}\orcidicon{0000-0003-3747-7120} \and
R.~Ottensamer\inst{\ref{inst:36}} \and 
I.~Pagano\inst{\ref{inst:12}}\orcidicon{0000-0001-9573-4928} \and
G. Peter\inst{\ref{inst:37}}\orcidicon{0000-0001-6101-2513} \and
D.~Piazza\inst{\ref{inst:4}} \and
G.~Piotto\inst{\ref{inst:34},\ref{inst:38}}\orcidicon{0000-0002-9937-6387} \and
D.~Pollacco\inst{\ref{inst:13}} \and 
D.~Queloz\inst{\ref{inst:39},\ref{inst:40}}\orcidicon{0000-0002-3012-0316} \and
R.~Ragazzoni\inst{\ref{inst:34},\ref{inst:38}}\orcidicon{0000-0002-7697-5555} \and
N. Rando\inst{\ref{inst:41}} \and 
H. Rauer\inst{\ref{inst:21},\ref{inst:42},\ref{inst:43}}\orcidicon{0000-0002-6510-1828} \and
I.~Ribas\inst{\ref{inst:15},\ref{inst:16}}\orcidicon{0000-0002-6689-0312} \and
G.~Scandariato\inst{\ref{inst:12}}\orcidicon{0000-0003-2029-0626} \and
D.~Ségransan\inst{\ref{inst:3}}\orcidicon{0000-0003-2355-8034} \and
A.~E.~Simon\inst{\ref{inst:4}}\orcidicon{0000-0001-9773-2600} \and
A. M. S. Smith\inst{\ref{inst:21}}\orcidicon{0000-0002-2386-4341} \and
M. Steller\inst{\ref{inst:11}}\orcidicon{0000-0003-2459-6155} \and
Gy.~M. Szabó\inst{\ref{inst:44},\ref{inst:45}} \and 
N.~Thomas\inst{\ref{inst:4}} \and 
S.~Udry\inst{\ref{inst:3}}\orcidicon{0000-0001-7576-6236} \and
V.~Van~Grootel\inst{\ref{inst:25}}\orcidicon{0000-0003-2144-4316} \and
N. A. Walton\inst{\ref{inst:46}}\orcidicon{0000-0003-3983-8778}
}

  \institute{
\label{inst:1} Instituto de Astrof\'isica de Canarias (IAC), E-38200 La Laguna, Tenerife, Spain \and
\label{inst:ull} Departamento de Astrof\'isica, Universidad de La Laguna (ULL), E-38206 La Laguna, Tenerife, Spain \and
\label{inst:2} Centre for Exoplanet Science, SUPA School of Physics and Astronomy, University of St Andrews, North Haugh, St Andrews KY16 9SS, UK \and 
\label{inst:3} Observatoire Astronomique de l'Université de Genève, Chemin Pegasi 51, Versoix, Switzerland \and
\label{inst:5} Center for Space and Habitability, University of Bern, Gesellschaftsstrasse 6, 3012 Bern, Switzerland \and
\label{inst:10} Dipartimento di Fisica, Universit\`a degli Studi di Torino, via Pietro Giuria 1, I-10125, Torino, Italy \and
\label{inst:4} Physikalisches Institut, University of Bern, Sidlerstrasse 5, 3012 Bern, Switzerland \and
\label{inst:6} Aix Marseille Univ, CNRS, CNES, LAM, 38 rue Frédéric Joliot-Curie, 13388 Marseille, France \and
\label{inst:7} Division Technique INSU, CS20330, 83507 La Seyne sur Mer cedex, France \and
\label{inst:13} Department of Physics, University of Warwick, Gibbet Hill Road, Coventry CV4 7AL, United Kingdom \and
\label{inst:8} Instituto de Astrofisica e Ciencias do Espaco, Universidade do Porto, CAUP, Rua das Estrelas, 4150-762 Porto, Portugal \and
\label{inst:9} Departamento de Fisica e Astronomia, Faculdade de Ciencias, Universidade do Porto, Rua do Campo Alegre, 4169-007 Porto, Portugal \and
\label{inst:11} Space Research Institute, Austrian Academy of Sciences, Schmiedlstrasse 6, A-8042 Graz, Austria \and
\label{inst:12} INAF, Osservatorio Astrofisico di Catania, Via S. Sofia 78, 95123 Catania, Italy \and
\label{inst:15} Institut de Ciencies de l'Espai (ICE, CSIC), Campus UAB, Can Magrans s/n, 08193 Bellaterra, Spain \and
\label{inst:16} Institut d'Estudis Espacials de Catalunya (IEEC), 08034 Barcelona, Spain \and
\label{inst:17} Admatis, 5. Kandó Kálmán Street, 3534 Miskolc, Hungary \and
\label{inst:18} Depto. de Astrofisica, Centro de Astrobiologia (CSIC-INTA), ESAC campus, 28692 Villanueva de la Cañada (Madrid), Spain \and
\label{inst:19} Université Grenoble Alpes, CNRS, IPAG, 38000 Grenoble, France \and
\label{inst:20} Department of Astronomy, Stockholm University, AlbaNova University Center, 10691 Stockholm, Sweden \and
\label{inst:21} Institute of Planetary Research, German Aerospace Center (DLR), Rutherfordstrasse 2, 12489 Berlin, Germany \and
\label{inst:22} Université de Paris, Institut de physique du globe de Paris, CNRS, F-75005 Paris, France \and
\label{inst:29} Science and Operations Department - Science Division (SCI-SC), Directorate of Science, European Space Agency (ESA), European Space Research and Technology Centre (ESTEC), Keplerlaan 1, 2201-AZ Noordwijk, The Netherlands \and
\label{inst:23} Centre for Mathematical Sciences, Lund University, Box 118, 221 00 Lund, Sweden \and
\label{inst:24} Astrobiology Research Unit, Université de Liège, Allée du 6 Août 19C, B-4000 Liège, Belgium \and
\label{inst:25} Space sciences, Technologies and Astrophysics Research (STAR) Institute, Université de Liège, Allée du 6 Août 19C, 4000 Liège, Belgium \and
\label{inst:34} INAF, Osservatorio Astronomico di Padova, Vicolo dell'Osservatorio 5, 35122 Padova, Italy \and
\label{inst:26} Leiden Observatory, University of Leiden, PO Box 9513, 2300 RA Leiden, The Netherlands \and
\label{inst:27} Department of Space, Earth and Environment, Chalmers University of Technology, Onsala Space Observatory, 439 92 Onsala, Sweden \and
\label{inst:28} University of Vienna, Department of Astrophysics, Türkenschanzstrasse 17, 1180 Vienna, Austria \and
\label{inst:30} Konkoly Observatory, Research Centre for Astronomy and Earth Sciences, 1121 Budapest, Konkoly Thege Miklós út 15-17, Hungary \and
\label{inst:31} ELTE E\"otv\"os Lor\'and University, Institute of Physics, P\'azm\'any P\'eter s\'et\'any 1/A, 1117 \and
\label{inst:37} Institute of Optical Sensor Systems, German Aerospace Center (DLR), Rutherfordstrasse 2, 12489 Berlin, Germany \and
\label{inst:32} IMCCE, UMR8028 CNRS, Observatoire de Paris, PSL Univ., Sorbonne Univ., 77 av. Denfert-Rochereau, 75014 Paris, France \and
\label{inst:33} Institut d'astrophysique de Paris, UMR7095, CNRS, Sorbonne Université, 98bis blvd. Arago, 75014 Paris, France \and
\label{inst:35} Astrophysics Group, Keele University, Staffordshire, ST5 5BG, United Kingdom \and
\label{inst:36} Department of Astrophysics, University of Vienna, Tuerkenschanzstrasse 17, 1180 Vienna, Austria \and
\label{inst:38} Dipartimento di Fisica e Astronomia "Galileo Galilei", Universita degli Studi di Padova, Vicolo dell'Osservatorio 3, 35122 Padova, Italy \and
\label{inst:39} ETH Zurich, Department of Physics, Wolfgang-Pauli-Strasse 2, CH-8093 Zurich, Switzerland \and
\label{inst:40} Cavendish Laboratory, JJ Thomson Avenue, Cambridge CB3 0HE, UK \and
\label{inst:41} ESTEC, European Space Agency, 2201AZ, Noordwijk, NL \and
\label{inst:42} Zentrum für Astronomie und Astrophysik, Technische Universität Berlin, Hardenbergstr. 36, D-10623 Berlin, Germany \and
\label{inst:43} Institut für Geologische Wissenschaften, Freie Universität Berlin, 12249 Berlin, Germany \and
\label{inst:44} ELTE E\"otv\"os Lor\'and University, Gothard Astrophysical Observatory, 9700 Szombathely, Szent Imre h. u. 112, Hungary \and
\label{inst:45} MTA-ELTE Exoplanet Research Group, 9700 Szombathely, Szent Imre h. u. 112, Hungary \and
\label{inst:46} Institute of Astronomy, University of Cambridge, Madingley Road, Cambridge, CB3 0HA, United Kingdom
}
  \date{Received September 15, 1996; accepted March 16, 1997}

  \abstract{
  Recent \tess-based studies have suggested that the dayside of \nplanet, a strongly-irradiated brown dwarf, is 
  significantly brighter in visible light than what would be expected based on \spitzer observations in infrared. 
  We observe eight eclipses of \nplanet with \cheops (CHaracterising ExOPlanet Satellite) to measure 
  its dayside brightness temperature in the bluest passband observed so far, and model the \cheops photometry 
  jointly with the existing optical and NIR photometry from \tess, LBT, CFHT, and \spitzer.
  Our modelling leads to a self-consistent dayside spectrum for \nplanet covering the \cheops, \tess, \pbh, \pbks, and \spitzer 
  IRAC 3.6 and 4.5 $\mu$m bands, where our \tess, \pbh, \pbks, and \spitzer band estimates largely agree 
  with the previous studies, but we discover a strong discrepancy between the \cheops and \tess bands. The \cheops 
  observations yield a higher photometric precision than the \tess observations, but do not show a significant eclipse signal, 
  while a deep eclipse is detected in the \tess band. The derived \tess geometric albedo of $0.36^{+0.12}_{-0.13}$ is difficult to 
  reconcile with a \cheops geometric albedo that is consistent with zero because the two passbands have considerable overlap. 
  Variability in cloud cover caused by the transport of transient nightside clouds to the dayside could provide an explanation 
  for reconciling the \tess and \cheops geometric albedos, but this hypothesis needs to be tested by future observations.}

   \maketitle

\section{Introduction}
\label{sec:introduction}

\nplanet is a strongly-irradiated transiting brown dwarf orbiting a 6500~K F5-star on a short-period orbit of 
1.2~days discovered by \citet{Siverd2012}. It has the third-shortest orbital period of the known transiting
brown dwarfs\footnote{According to the \href{https://www.theroncarmichael.com/population}{Transiting Brown Dwarfs} 
list accessed 16.2.2022.} \citep{Carmichael2021} after TOI-263b \citep[$P=0.56$~d,][]{Parviainen2020,Palle2021} and NGTS-7Ab 
\citep[$P=0.7$~d,][]{Jackman2019}, but receives a significantly higher stellar insolation since TOI-263b and NGTS-7Ab 
both orbit low-mass M~dwarfs with $\teff\sim3400$~K. The orbit and insolation level of \nplanet is similar 
to ultra-hot Jupiters \citep{Parmentier2018,Kitzmann2018,Lothringer2018} like MASCARA-2b \citep{Lund2017,Talens2018}, 
WASP-121b \citep{Delrez2015}, WASP-189b \citep{Anderson2018}, WASP-76b \citep{West2016}, and WASP-33b \citep{Cameron2010}, but its surface gravity 
is $\sim$22 times larger than that of Jupiter. This makes \nplanet an interesting case to study how surface gravity
affects the atmospheres of strongly-irradiated sub-stellar objects.

\nplanet's emission spectrum\footnote{e use a very loose definition of "emission spectrum" that also includes
the contribution from reflected light.} has been studied using ground- and space-based eclipse and phase curve
observations covering wavelengths from red-visible (\tess) to 4.5~$\mu$m (\spitzer). Soon after \nplanet's discovery, 
\citet{Beatty2013} observed one secondary eclipse using \spitzer 
(in IRAC 3.6 and 4.5~$\mu$m passbands) and four eclipses in $z'$-band using the 0.6~m RCOS telescope at 
Moore Observatory. Later, \citet{Croll2015} observed an eclipse in the \pbks-band using the WIRCam on the Canada-France-Hawaii 
Telescope, \citet{Beatty2017} observed an eclipse in the \pbh-band using the LUCI1 spectrograph on the Large Binocular 
Telescope, and \citet{Beatty2019} observed a full \nplanet phase curve with \spitzer. Finally, the \tess telescope 
observed \nplanet continuously for 25 days in 2019, and the photometry spanning 15 full orbits was studied by 
\citet{Beatty2020} and \citet{VonEssen2021}.

The previous eclipse and phase curve studies have shown that \nplanet's eclipse spectrum in near- and mid-IR agrees
with a constant dayside brightness temperature of $\sim$2900~K \citep{Beatty2019}. However, a study of the \tess
observations by \citet{Beatty2020} presented that the brown dwarf's dayside brightness temperature in the \tess band 
is significantly higher than in the \spitzer bands. They proposed that this higher-than-expected brightness 
in visible light could be due to a high albedo caused by silicate clouds forming in the nightside of the brown dwarf. 
\nplanet's dayside is too hot for clouds to form, but the nightside temperature is expected to be cool enough 
for the formation of silicate clouds. These nightside clouds could then be blown over to the dayside by winds,
where they could survive all the way to the local noon, boosting the otherwise low dayside albedo significantly.

However, the cloud hypothesis is not the only viable explanation for the unexpected dayside brightness in the \tess 
band. \citealt{VonEssen2021} made an independent analysis of
the \tess phase curve and proposed that \nplanet's dayside emission spectrum can be explained with thermal emission
alone. \citet{VonEssen2021} suggest that the differences in the \tess and \spitzer brightness temperatures
can be explained by collision-induced absorption due to H$_2$-H$_2$ and H$_2$-He decreasing the brightness
temperature in the \spitzer bands. So, rather than the brightness of the \tess passband being boosted by 
reflection from clouds, the \spitzer band brightness would be suppressed by molecular absorption.

\cheops \citep[\textit{CHaracterising ExOPlanet Satellite};][]{Benz2021} is an ESA S-class mission aiming
to characterise transiting exoplanets around bright ($V<12$) stars based on ultra-high quality photometry
\citep{Broeg2013}. The satellite is equipped with a 30~cm telescope that allows it to reach a photometric
precision of 20~ppm in 6~h of integration time for a $V=9$ star (as a design requirement), which is sufficient 
to detect Earth-sized planets around G5 dwarf stars \citep{Benz2021}. 
\cheops was launched on 18 December 2019, the science observations started in April 2020, and the observations
have already been used to improve the characterisation of new and known exoplanet systems
\citep{Bonfanti2021,Delrez2021,Szabo2021,Barros2022,Lacedelli2022, Wilson2022},
improve the characterisation of eclipsing binaries with M-dwarf companions \citep{Swayne2021},
search for transit timing variations \citep[TTVs, ][]{Borsato2021}, search for planets transiting white
dwarfs \citep{Morris2021a}, and study secondary eclipses and full phase curves of super-Earths and hot 
Jupiters \citep{Lendl2020,Morris2021,Hooton2022,Deline2022}. By now, the satellite has shown it can
surpass its design requirements by reaching 10-20~ppm photometric precision for one hour of integration
\citep{Lendl2020}.

In this paper, we present the phase curve analysis of \nplanet based on new \cheops secondary eclipse 
photometry and existing photometry from \tess, \spitzer, LUCI1, and WIRCam. While the \cheops and \tess
passbands have a significant overlap, the \cheops band emphasises bluer wavelengths than the \tess band.
This difference allows us to test the two hypotheses about the \tess-\spitzer dayside brightness temperature
discrepancy. If the high-albedo hypothesis is correct, both the \cheops- and \tess-bands are expected to 
be dominated by reflected light, and the \cheops-band can be expected to also show a high albedo (albeit
this cannot be taken for granted since there are scenarios where \nplanet could have a low albedo in
the \cheops band and high in the \tess-band, as will be discussed later). If the IR absorption hypothesis 
is correct, \cheops passband brightness temperature could be expected to follow closely the model presented 
in Fig.~9 in \citet{VonEssen2021}.

All the data and code for the analysis and figures are publicly available from \ghlink{}{GitHub}, including  
additional Jupyter notebooks that work as appendices to the paper.

\section{Observations}
\label{sec:observations}
\subsection{CHEOPS}
\label{sec:observations.cheops}

\begin{figure*}
	\centering
	\includegraphics[width=\textwidth]{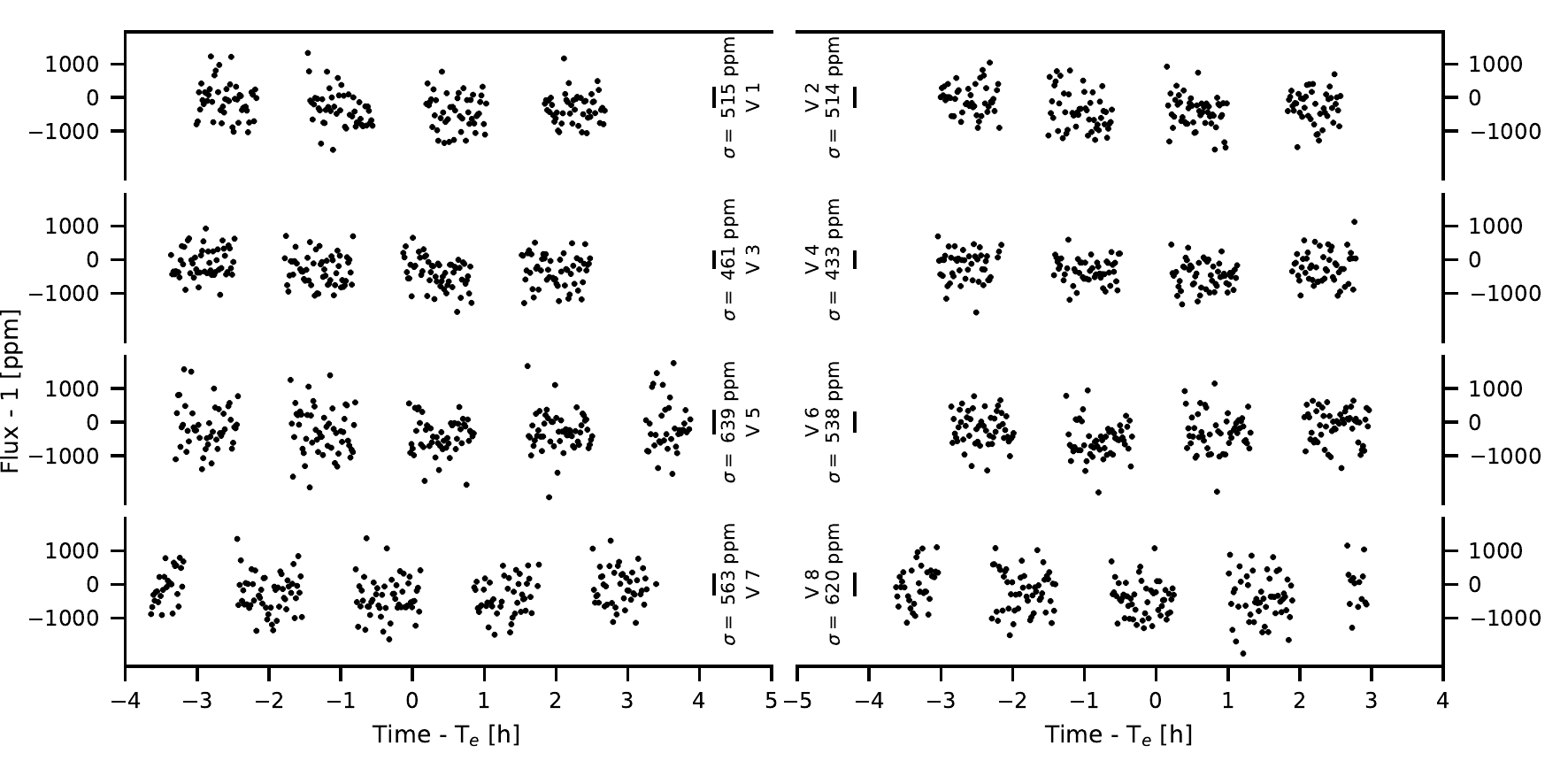}
	\caption{The detrended photometry from eight \nplanet eclipses observed with \cheops centred around the eclipse centre time ($T_e$). 
	Each eclipse is observed as a single continuous \cheops visit of $\approx$7~h (labelled as V1-V8 in the figure), but the 
	satellite's orbital configuration leads to periodic gaps in the photometry, as described in Sect.~\ref{sec:observations.cheops}.}
	\label{fig:cheops_photometry}
\end{figure*}

We observed eight eclipses of \nplanet with the \cheops spacecraft between 2020-10-02 and 2020-12-01 as a part of the Guaranteed Time 
Observers (GTO) programme (Fig.~\ref{fig:cheops_photometry}). 
Each eclipse observation consisted a \cheops visit of $\approx$7~h of near-continuous observations with an exposure time of 60~s. 
The length of the visits combined with the high precision of the \nplanet ephemeris (thanks to recent \tess observations) ensured that 
we obtained at least 2~h of pre- and post-eclipse baseline for each visit.
As \cheops is in a low-Earth orbit, sections of the observations are unobtainable as they may 
occur during Earth-target occultations, passages through the South Atlantic Anomaly (SAA), or at times when the stray light level 
passes an acceptable threshold. These effects manifest as gaps in the light curves and result in lower efficiency observations. For 
the data presented here, we obtain 
an average efficiency of 56\%.

The data were automatically processed using the latest version of the \cheops Data Reduction Pipeline (DRP v13; \citealt{Hoyer2020}) 
that performs image calibration including bias, gain, non-linearity, dark current, and flat fielding corrections, and amends of 
environmental and instrumental effects, for example cosmic-ray hits, smearing trails, and background variations. Subsequently, 
aperture photometry is conducted on the corrected images using a set of standard apertures; $R$ = 22.5\arcsec (RINF), 25.0\arcsec 
(DEFAULT), and 30.0\arcsec (RSUP), with an additional aperture that selects the radius based on contamination level and instrumental 
noise (OPTIMAL), which for these observations yields 20.5\arcsec. Furthermore, the DRP calculates contamination estimates of the visits 
that comes from background sources, as detailed in Section~6.1 of \cite{Hoyer2020}, which we remove from the data.

As the field of view of \cheops rotates due to the nadir-locked orbit of the spacecraft, there may be short-term, non-astrophysical 
flux trends on the order of a \cheops orbital timescale due to nearby contaminants, background variations, or changes in instrumental 
environment. In previous studies these trends have been corrected via linear decorrelation with instrumental basis vectors
\citep{Bonfanti2021,Delrez2021} or Gaussian process regression \citep{Lendl2020}, however a recent study has identified that these 
roll angle trends can be corrected by a novel PSF detrending method \citep{Wilson2022}. To summarise, this tool determines PSF shape 
changes of a visit by computing vectors associated with these variations by conducting principal component analyses (PCA) of the 
auto-correlation functions of the \cheops subarray images. The components that contribute most to the PSF shape changes are subsequently 
selected by a leave-one-out-cross-validation and are used as covariates in the light curve model. For this study,
we perform this method on all visits using the DEFAULT aperture photometry fluxes as these yield the lowest RMS (root mean square) noises.

\subsection{TESS}
\label{sec:observations.tess}
The Transiting Exoplanet Survey Satellite \citep[\tess,][]{Ricker2014} observed \nplanet during Sector 17 for 25 days (2019-10-08--2019-11-02) with a two-minute cadence.
The observations cover 15 orbital periods with a high-enough precision to measure all the significant phase curve components in the \tess passband. 

We use the Presearch Data Conditioning (PDC-SAP) light curve \citep{Stumpe2014,Stumpe2012,Smith2012a} produced by the SPOC pipeline \citep{Jenkins2016}.
As in \citet{Beatty2020}, we remove all the exposures with a non-zero quality flag, but we do not apply any additional detrending or outlier 
rejection. Instead, we use a \celerite-based \citep{Foreman-Mackey2017} time-dependent Gaussian process (GP) to calculate the likelihood as explained in more detail later. 
Furthermore, we ignore the residual light contamination of $\approx0.82\%$ reported by \citet{VonEssen2021}, since it will not have any practical
effect on the results.

\subsection{Spitzer}
\label{sec:observations.spitzer}

\nstar was observed with the \spitzer space telescope using the Infrared Array Camera (IRAC; \citealt{Fazio2004}) in both \textit{S1} 
(3.6~$\mu$m) and \textit{S2} (4.5~$\mu$m) passbands across multiple programs with the observations covering a secondary eclipse 
\citep{Beatty2017} and a full phase curve \citep{Beatty2019}.

Using these data, we extract light curves for all visits using our own custom pipeline. Using the flux-calibrated and artefact-corrected CBCD 
(Corrected Basic Calibrated Data) frames we constructed light curves by firstly using the IRAC archive contributed IDL program {\sc box\_centroider.pro}\footnote{\url{https://irsa.ipac.caltech.edu/data/SPITZER/docs/irac/calibrationfiles/pixelphase/box_centroider.pro}}
to locate the target in the images. In addition to utilising these values as decorrelation basis vectors during the joint fitting, these 
x- and y-centroid positions were used as inputs into the {\sc aper.pro} code to perform aperture photometry using an aperture radius of 3\,pixels
and a sky annulus 12--20\,pixels. Lastly, we perform the aperture corrections, correct intrapixel gain using the {\sc iracpc\_pmap\_corr.pro}\footnote{\url{https://irsa.ipac.caltech.edu/data/SPITZER/docs/dataanalysistools/tools/contributed/irac/iracpc_pmap_corr/}} tool, 
and reject data with fluxes or x- and y-centroid positions 5$\sigma$ away from the respective, normalised mean.

\subsection{Large Binocular Telescope H-band}
\label{sec:observations.lbt}

\citet{Beatty2017} observed a spectrally resolved eclipse of \nplanet in the \pbh-band using the LUCI1 multiobject spectrograph on 
the 2$\times$8.4~m Large Binocular Telescope (LBT). The observations were carried out on the night of 2013-10-26 and lasted $\approx$5~h. 
The exposure time was 60~s, leading to 245 exposures in total.\!\footnote{The LBT \pbh-band light curves used in \citet{Beatty2017} were
kindly provided by Dr.~Beatty, personal communication.}

\subsection{Canada-France-Hawaii Telescope \pbks-band}
\label{sec:observations.cfht}

\citet{Croll2015} observed an eclipse of \nplanet in the \pbks-band ($\sim$2.15~$\mu$m) using the Wide-field Infrared Camera (WIRCam) 
on the 3.6~m Canada-France-Hawaii Telescope (CFHT). The observations were carried out in 2012-10-10 and covered 6.8~h centred around 
the expected eclipse centre with an exposure time of 8~s, leading to 1440 exposures in total.\!\footnote{The CFHT \pbks-band 
light curve used in \citet{Croll2015} was kindly provided by Dr.~Croll, personal communication.} 
The reduction of the photometry is detailed extensively in \citet[][see especially the discussion about the sensitivity to
different data reduction approaches in Sect.~5.4]{Croll2015}.

\section{Stellar Characterisation}
\label{sec:stellar_characterisation}

We revise the stellar radius, mass, and age and show the updated stellar properties in Table~\ref{table:star}. The stellar age and
mass are compatible with the previous estimates by \citet{Siverd2012} within $1\sigma$, but our new radius is slightly larger than
the original estimate of $1.46^{+0.039}_{-0.030}\;\rsun$. 

Using the stellar parameters derived via our spectral analysis as priors, we use a Markov-Chain Monte Carlo (MCMC) modified infrared flux method 
(IRFM; \citealt{Blackwell1977,Schanche2020}) to compute the stellar radius of \nstar. We construct spectral energy distributions (SEDs) using 
\textsc{atlas} Catalogue stellar atmospheric models \citep{Castelli2003} and subsequently compute synthetic fluxes by integrating the SEDs over 
passbands of interest with attenuation of the SED included as a free parameter to account for extinction. To calculate the stellar apparent
bolometric flux, hence the stellar angular diameter and effective temperature, we compare the synthetic fluxes to the observed broadband photometry 
retrieved from the most recent data releases for the following passbands; {\it Gaia} G, G$_{\rm BP}$, and G$_{\rm RP}$, 2MASS J, H, and K, and 
{\it WISE} W1 and W2 \citep{GaiaCollaboration2021,Skrutskie2006,Wright2010}. We convert the angular diameter into stellar radius using 
the offset-corrected {\it Gaia} EDR3 parallax \citep{Lindegren2021} and determine a $R_{\star}=1.530\pm0.009\,R_{\odot}$.

We then use \teff, [Fe/H], and $R_{\star}$ to compute the stellar mass $M_{\star}$ and age $t_{\star}$ through two different sets of theoretical 
evolutionary models. In detail, we retrieved a first pair of mass and age estimates by applying the isochrone placement technique 
\citep{bonfanti15,bonfanti16} to grids of stellar tracks and isochrones pre-computed by the PARSEC\footnote{\textit{PA}dova and T\textit{R}ieste 
\textit{S}tellar \textit{E}volutionary \textit{C}ode: \url{http://stev.oapd.inaf.it/cgi-bin/cmd}} v1.2S code \citep{marigo17}. A second pair of 
mass and age estimates, instead, was derived by directly inputting \teff, [Fe/H], and $R_{\star}$ into the CLES\footnote{Code Liègeois d'Évolution 
Stellaire} code \citep{scuflaire08}, which computes the best-fit evolutionary track following the Levenberg-Marquadt minimisation scheme, as presented 
in \citet{salmon21}. As described in \citet{Bonfanti2021}, once the two pairs of outcomes are available, we checked their mutual consistency through 
a $\chi^2$-based criterion and finally merged the posterior distributions obtaining $M_{\star}=1.370_{-0.070}^{+0.047}\,M_{\odot}$ and 
$t_{\star}=1.8\pm0.5$~Gyr.

\begin{table}[t]    
	\caption{\nstar identifiers, coordinates, properties, and magnitudes.}
	\centering
	\begin{tabular*}{\columnwidth}{@{\extracolsep{\fill}} llrrr}
		\toprule\toprule
		\multicolumn{5}{l}{\emph{Main identifiers}}     \\
		\midrule 
		TIC      & \multicolumn{4}{r}{432549364} \\
		GAIA DR2 & \multicolumn{4}{r}{2881784280929695360} \\
		2MASS   & \multicolumn{4}{r}{J00012691+3923017}   \\ 
		\\
		\multicolumn{5}{l}{\emph{Equatorial coordinates}}     \\
		\midrule            
		RA \,(J2000) & \multicolumn{4}{r}{$00^h\,01^m\,26\fs9$}            \\
		Dec (J2000)  & \multicolumn{4}{r}{ $39\degr\,23\arcmin\,01\farcs66$}  \\
		\\     
		\multicolumn{5}{l}{\emph{Stellar parameters }} \\
		\midrule
		Spectral type & & & F5 & (1)\\
		Mass             & $M_\star$ & [\msun]  & $1.370_{-0.070}^{+0.047}$ & (2) \\
		Radius           & $R_\star$ & [\rsun]  & $1.530 \pm 0.009$    & (2) \\
		Density          & $\rho$    &  [\gcm]  & $0.539 \pm 0.025$    & (2) \\
		Eff. temperature & \teff     & [K]      & $6516 \pm 49$        & (1) \\
		Surf. gravity    & $\log g$  & [cgs]    & $4.23 \pm 0.02$      & (1) \\
		Metallicity      & [Fe/H]    & [dex]    & $0.05 \pm 0.08$      & (1) \\
		Age              &           & [Gyr]    & $1.8\pm0.5$          & (2) \\
		Proj. rot. velocity & $v \sin i$ & [km\,s$^{-1}$] & $56 \pm 2$ & (1) \\
		Parallax         &           & [mas]    & 3.684\,$\pm$\,0.014  & (3) \\
		\\
		\multicolumn{5}{l}{\emph{Magnitudes}} \\
		\midrule              
		\centering
		
		Filter & \multicolumn{2}{r}{Magnitude}       & Uncertainty & \\
		\midrule     
		TESS & & 10.2242 & 0.0061 &\\
		$B$  & & 11.233 & 0.11 &\\
		$V$  & & 10.632 & 0.007 &\\
		Gaia & & 10.5905 & 0.00034 &\\
		$J$  & & 9.682 & 0.022 &\\
		$H$  & & 9.386 & 0.030 &\\
		$K$  & & 9.437 & 0.019 &\\
		\bottomrule
	\end{tabular*}
	\tablebib{(1)~\citet{Siverd2012}; (2)~This work; Gaia EDR3 \citep{GaiaCollaboration2021}.}
	\label{table:star}  
\end{table}

\section{Theory: phase curve modelling}
\label{sec:theory}

\begin{figure}
	\centering
	\includegraphics[width=\columnwidth]{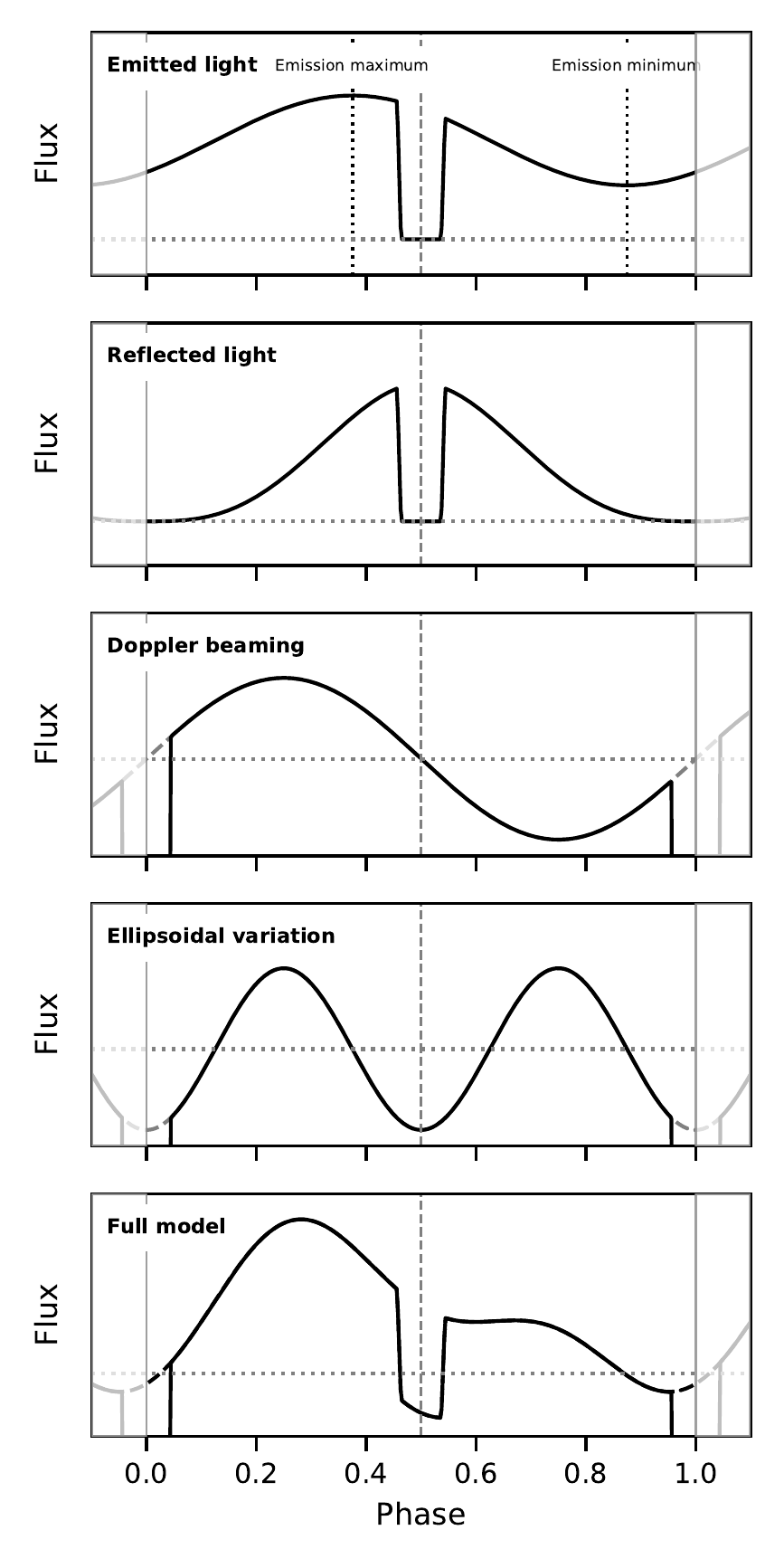}
	\caption{Components of a phase curve model (transit, eclipse, emitted light, reflected light, Doppler beaming, 
	and ellipsoidal variation) and the full model. The figure is centred around the secondary eclipse, the 
	horizontal dotted line shows the baseline level (flux = 1), the vertical dashed line shows the eclipse centre, and the
	solid grey line shows the transit centre.}
	\label{fig:phase_curve_components}
\end{figure}

\subsection{Phase curve model}
\label{sec:theory.model}

We model the light curves using a phase curve model implemented in \pytransit v2.5.21 
\citep{Parviainen2015,Parviainen2020a,Parviainen2020b} that includes a primary transit, secondary eclipse, thermal emission, 
reflection, Doppler beaming, and ellipsoidal variation, as illustrated in Fig.~\ref{fig:phase_curve_components}. 
The observed flux is presented as a sum of stellar and planetary components divided by a baseline flux model as 
\begin{equation}
    F = \frac{T \times \left( f_\mathrm{EV} + f_\mathrm{DB} \right) + E \times k^2 (f_\mathrm{R} + f_\mathrm{E})}{f_b}, \label{eq:flux_model}
\end{equation}
where $f_b$ represents the baseline flux, $T$ is the transit model, $f_\mathrm{EV}$ is the ellipsoidal variation component, 
$f_\mathrm{DB}$ is the Doppler beaming component, $E$ is the eclipse model, $k$ is the planet-star radius ratio, 
$f_\mathrm{R}$ is the reflected flux, and $f_\mathrm{E}$ is the thermal emission.\!\footnote{Note that the $f_\mathrm{R}$ and $f_\mathrm{E}$ represent
the phase-dependent planet-star \emph{flux ratios} from reflection and thermal emission, where the \emph{flux ratio} stands for the
ratio of the radiation emitted towards the observer by the brown dwarf and the star per projected unit area. That is, the \emph{flux ratio}
is not the ratio of the disk-integrated fluxes, but this can be obtained by multiplying the flux ratio by the companion-star area ratio, $k^2$.} 

\begin{table}
    \centering
    \caption{Phase curve model parameters and priors used in the final joint analysis. The \textit{global} 
    parameters are independent of passband or light curve, the \textit{passband-dependent} parameters are 
    repeated for each passband, and the \textit{light-curve-dependent} parameters are repeated for each separate 
    light curve. N$(\mu, \sigma)$ stands for a normal prior with a mean $\mu$ and standard deviation $\sigma$,
    and U$(a,b)$ stands for a uniform distribution from $a$ to $b$.}
    \label{table:model_parameters}
    \begin{tabular*}{\columnwidth}{@{\extracolsep{\fill}} llll}
	\toprule\toprule
    Description & Parameter & Units & Prior \\
	\midrule     
	\multicolumn{4}{l}{\emph{Global parameters}} \\
	\midrule
    Mid-transit time & $t_0$  & d & N($T_\mathrm{S12}$)$^a$\\
    Orbital period   & $p$    & d & N($P_\mathrm{S12}$)$^a$\\
	Area ratio & $k^2$  & - & N(0.078, 0.005) \\
    Stellar density  & $\rho$ & \gcm & N(0.54, 0.08) \\
    Impact parameter & $b$    & - & U(0, 1) \\
                     & $\sqrt{e} \cos \omega$ & - & N(0, $10^{-6}$)\\
                     & $\sqrt{e} \sin \omega$ & - & N(0, $10^{-6}$) \\
	EV offset & $o_\mathrm{EV}$ & deg & N(0, 5) \\
    \\
	\multicolumn{4}{l}{\emph{Passband-dependent parameters}} \\
	\midrule
	Quad. LD coeff. 1 & $q_1$ & - & \ldtk$^b$ \\
	Quad. LD coeff. 2 & $q_2$ & - & \ldtk$^b$ \\
	EV amplitude & $A_\mathrm{EV}$  & - & U$^c$ \\
	DB amplitude & $A_\mathrm{DB}$  & - & N$^d$ \\
	Emission offset & $o_\mathrm{e}$ & deg & N(0, 10) \\
	Log$_{10}$ ds. emission & $\log_{10} f_\mathrm{d}$  & - & U(-3, 0)\\
	Log$_{10}$ ns. emission & $\log_{10} f_\mathrm{n}$ & - & U(-3, 0) \\
    Geometric albedo & $a_\mathrm{g}$ & - & N($10^{-5}$, $10^{-7}$) \\
    \\
    \multicolumn{4}{l}{\emph{Light-curve-dependent parameters}} \\
	\midrule
	Baseline level & $f_0$ & - & N$^e$ \\
	Log$_{10}$ white noise & $\log_{10} \sigma$ & - & N$^f$ \\
	Ln GP input scale & $s_i$ & - & N(1, 1)$^g$ \\
	Ln GP output scale & $s_o$ & - & N$^g$ \\
    \bottomrule
    \end{tabular*}
    \tablefoot{
		\tablefoottext{a}{The zero epoch and orbital period priors are based on the values reported in \citet{Siverd2012} with uncertainties
		multiplied by three.}
		\tablefoottext{b}{The limb darkening coefficients have normal priors calculated using \ldtk.}
		\tablefoottext{c}{Ellipsoidal variation is constrained based by EV amplitude ratios as explained in the text and listed in Table~\ref{table:ev_and_db_priors}}
		\tablefoottext{d}{The Doppler beaming priors are given in Table~\ref{table:ev_and_db_priors}.}
		\tablefoottext{e}{The baseline levels have wide normal priors that are based on the light curve variability.}
		\tablefoottext{f}{The average log white noise parameters have normal priors centred around a numerical white noise estimate.}
		\tablefoottext{g}{The Gaussian process log input scales have wide normal priors while the log output scales have normal priors based on light curve variability.}
		}
\end{table}

The model parametrisation is shown in Table~\ref{table:model_parameters} and the parameter priors are discussed in more detail
in Sect.~\ref{sec:analysis.priors}. We assume a circular orbit since \nplanet's orbital eccentricity agrees with zero 
\citep{Siverd2012}, which allows us to model the phase curve simply as a function of the orbital phase, $\phi = 2\pi (t-t_0) / p$, 
where $t$ is the mid-exposure time, $t_0$ is the mid-transit time, and $p$ is the orbital period.

The light curves are modelled jointly, and each light curve has its own baseline and noise model. We use a simple constant
baseline model (that is, a normalisation factor), and model the systematics as a Gaussian process, as detailed
later. All the baseline and noise model parameters are free (or loosely constrained) in the posterior optimisation and sampling.

\subsection{Transits and eclipses}
\label{sec:theory.transits_and_eclipses}
The transits are modelled using the quadratic limb darkening model by \citet{Mandel2002} as implemented in \pytransit 
\citep{Parviainen2015}, and the limb darkening is parameterised using the triangular parameterisation by
\citet{Kipping2013b}. The secondary eclipses are also modelled using \pytransit, using an eclipse model that returns
the fraction of visible planetary disk area to the stellar disk area. We do not include light travel time into the 
eclipse modelling since it is at maximum $\approx 25$~s.

\subsection{Reflected light}
\label{sec:theory.reflection}
The reflected light\footnote{We fix the reflected light component to zero for the main analyses due to its degeneracy 
with the thermal emission, but describe the component here for consistency.} 
is approximated using a Lambertian phase function \citep{Russell1916,Madhusudhan2011}. 
This choice is based on simplicity since our data is not precise enough to justify a more realistic (and flexible) 
phase function. The planet to star flux ratio from reflected light is
\begin{equation}
    f_\mathrm{R} = \frac{a_\mathrm{g}}{a_s^2} \frac{\sin\alpha + (\pi - \alpha) \cos\alpha}{\pi}.
\end{equation}
where $a_g$ is the geometric albedo, $a_\mathrm{s}$ the scaled semi-major axis, $a_\mathrm{s} = a/R_\star$, 
and $\alpha$ is the phase angle, $\alpha = |\phi - \pi|$. The reflected light reaches its maximum near the
eclipse and goes to zero near the transit.

\subsection{Thermal emission}
\label{sec:theory.emission}
The thermal emission from the companion is modelled as a sine wave between the minimum flux ratio, 
$f_\mathrm{n}$, and maximum flux ratio, $f_\mathrm{d}$, as
\begin{equation}
    f_\mathrm{E} = f_\mathrm{n} + \frac{f_\mathrm{max} - f_\mathrm{min}}{2}\left(1-\cos(\phi + o)\right),
\end{equation}
where $o$ is a phase offset that sets the location of the hot spot, moving the minimum and maximum
emission phases marked in Fig.~\ref{fig:phase_curve_components}.

\subsection{Ellipsoidal variation}
\label{sec:theory.ev}
When a star is orbited by a massive companion on a short-period orbit, the companion's gravitational pull 
distorts the star's shape from a sphere into an ellipse. This causes an ellipsoidal variation (EV) signal 
\citep{Morris1985,Faigler2011,Barclay2012,Lillo-Box2014} that can be approximated to a first order as
\begin{equation}
    f_\mathrm{EV} = -A_\mathrm{EV}\cos(2\phi + o),
\end{equation}
where $o$ is an angular offset, the amplitude, $A_\mathrm{EV}$, depends on the companion-host mass ratio,
orbital inclination, $i$, and scaled semi-major axis, $a_\mathrm{s}$, as
\begin{equation}
    A_\mathrm{EV} = \beta \frac{M_\mathrm{c}}{a_\mathrm{s}^3 M_\star} \sin^2(i),
\end{equation}
where $\beta$ is a factor that depends on the linear limb darkening coefficient, $u$, and gravity darkening coefficient, $g$, as
\begin{equation}
    \beta = 0.15 \frac{(15 + u)(1 + g)}{3 - u}. \label{eq:ev_beta}
\end{equation}

The EV signal is usually inconsequential when studying planetary phase curves, but it can dominate over the
other phase curve components when a star is orbited by a massive companion on a short-period orbit, such as in 
the case of \nplanet. Including the EV component into an eclipse and phase curve model is especially important
because the EV signal has its minimum close to the mid-eclipse time \citep[][and Sect.~6.3 in 
\citealt{VonEssen2021}]{Bell2019,Beatty2020}. Unaccounted for EV signal can artificially 
enhance the eclipse depth estimated from secondary eclipse observations, as happened with some of the previous
\nplanet observations.

\subsection{Doppler beaming}
\label{sec:theory.beaming}
Doppler beaming is calculated following \citet{Loeb2003}, \citet{Barclay2012}, and \citep{Claret2020} as
\begin{equation}
    f_\mathrm{DB} = A_\mathrm{DB} \sin \phi,
\end{equation}
where the Doppler beaming amplitude is
\begin{equation}
    A_\mathrm{DB} =  \frac{\hat{B}}{c} \left( \frac{2\pi G}{p}\right)^{1/3} \frac{M_p \sin i}{M_\star^{2/3}},
\end{equation}
and $c$ is the speed of light in vacuum, $G$ is the gravitation constant, $p$ is the orbital period, and $\hat{B}$ 
is the photon-weighted passband-integrated beaming factor \citep{Bloemen2010}. The passband-integrated beaming factor 
can be calculated from a stellar spectrum for any given passband as 
 \begin{equation}
  \hat{B} = \frac{\int B\,\lambda\,F(\lambda)T(\lambda)\,\ud\lambda}{\int \lambda\,F(\lambda)T(\lambda)\,\ud\lambda},
\end{equation}
where $T(\lambda)$ is the passband transmission, $F_\lambda$ is the stellar flux at wavelength $\lambda$, and 
$B = 5 + \ud \log F_\lambda / \ud \log \lambda$ is the beaming factor.

Unlike ellipsoidal variations, Doppler beaming has only a minimal effect on the observed eclipse depth. This is
because the beaming signal behaves as a linear slope at the vicinity of the eclipse. In addition, beaming is 
expected to have a much lower amplitude for \nplanet than EV.

\section{Phase curve analysis}
\label{sec:analysis}
\subsection{Overview}

The goal of our study is to estimate the day- and nightside flux ratios between \nplanet and its host star 
in the \cheops, \tess, \pbh, \pbks, and \spitzer 3.6 and 4.5~$\mu$m passbands. We do this by modelling the
light curves described in Sect.~\ref{sec:observations} jointly using the phase curve model defined 
in Sect.~\ref{sec:theory}, which will also yield improved orbital and geometric parameter estimates as
a by-product. The analysis follows the standard procedures of Bayesian inference
\citep{Gelman2013,Parviainen2018}, where we aim to estimate the posterior probability distributions 
(posteriors) for the model parameters given a model, observations, and prior probability distributions 
(priors) for the model parameters.

We carry out three analyses that are each further divided into separate scenarios (summarised in Appendix 
\ref{sec:analysis_scenarios}) with different priors on the phase
curve model parameters. The analyses are:
\begin{enumerate}
    \item \textbf{External data analysis} models the \tess, LBT, CFHT, and \spitzer observations jointly, and 
    is implemented in \lpfed Python class. The external data analysis was carried out to create priors for the 
    \cheops analysis, and also allows us to test how including the \cheops observations affects the final 
    parameter estimates. 
    \item \textbf{\cheops analysis} models only the \cheops-observed eclipses, and is implemented in the 
    \ghlink{src/cheopslpf.py}{\texttt{CHEOPSLPF}} class. The main motivation for the \cheops analysis is to
    create detrended \cheops light curves to be used in the final analysis. As mentioned in 
    Sect.~\ref{sec:observations.cheops}, the \cheops light curves contain strong instrument-related systematics
    that are captured as basis vectors (covariates) using the approach described in \citet{Wilson2022}.
    Modelling the \cheops observations jointly using a linear baseline model leads to a model with 180
    free parameters, most of which are baseline model coefficients. This is a lot considering that the external
    data analysis has 79 free parameters, most of which are physically interesting. After some tests, 
    we decided to carry out the \cheops analysis with a full baseline model and simplify the final analysis
    by using the \cheops analysis posterior baseline model to remove the systematics from the \cheops 
    light curves included in the final analysis.
    \item \textbf{Final analysis} models all the data jointly, and is implemented in the \lpffn class.
\end{enumerate}
The posterior estimates from the final analysis are adopted as the final results, but the differences between the
analyses and scenarios are discussed in Appendix~\ref{sec:analysis_scenarios}.

All the analyses are implemented as Python classes that inherit \texttt{pytransit.lpf.PhaseCurveLPF}, where the base class 
implements the functionality to model phase curves jointly for a set of heterogeneous light curves\footnote{Heterogeneous in 
the sense that they may have been observed in different passbands, exposure times (some light curves possibly needing model 
supersampling and others not), and each possibly with their own noise and systematics models.} (including optimisation and 
Markov Chain Monte Carlo sampling), and the derived classes mainly read in the different light curves, set up the noise
models, and set up the model parameter priors based on the analysis scenario.

\subsection{Priors}
\label{sec:analysis.priors}

\subsubsection{Orbit, geometry, and limb darkening}
\label{sec:analysis.priors.general}

For the final and external dataset analyses, we set uninformative or only slightly informative priors on the geometric 
and orbital parameters, such as the orbital period, impact parameter, and companion-star radius ratio, as listed in 
Table~\ref{table:model_parameters}. We also assume that the radius ratio is constant from passband to passband because
of a small atmospheric scale height caused by \nplanet's high surface gravity \citep{Beatty2020}. Further, we assume a 
circular orbit and force eccentricity to zero, since both the RV and eclipse observations agree with a circular orbit 
\citep{Siverd2012}. 

The transit model adopts a quadratic limb darkening law with the \citet{Kipping2013b} parametrisation. We set uninformative 
priors on the \tess band limb darkening coefficients since the \tess light curve does not feature any strong systematics and 
the transit SNR is high. However, we set \ldtk-calculated normal priors on the limb darkening  coefficients for the two 
\spitzer passbands because the strong systematics in the \spitzer light curves will not allow us to constrain limb darkening 
well.

\subsubsection{Doppler beaming}
\label{sec:analysis.priors.db}

\begin{table*}
    \centering
    \caption{Theoretical Doppler beaming and ellipsoidal variation amplitudes, and ellipsoidal variation amplitude 
    ratios relative to the \tess passband.}
    \label{table:ev_and_db_priors}
    \begin{tabular*}{\textwidth}{@{\extracolsep{\fill}} lrrr}
	\toprule\toprule
    Passband & Doppler beaming amplitude [ppm] & EV amplitude [ppm] & EV amplitude ratio \\
	\midrule     
    \cheops             & $63 \pm 3.2$ & $479 \pm 30$ & $1.083 \pm 0.007$ \\
    \tess               & $41 \pm 2.1$ & $442 \pm 28$ & 1.00 \\
    \pbh                & $25 \pm 1.3$ & $358 \pm 23$ & $0.809 \pm 0.004$ \\
    \pbks               & $19 \pm 9.2$ & $352 \pm 22$ & $0.796 \pm 0.004$ \\
    \spitzer 3.6~$\mu$m & $17 \pm 8.3$ & $343 \pm 22$ & $0.774 \pm 0.003$ \\
    \spitzer 4.5~$\mu$m & $16 \pm 7.9$ & $341 \pm 22$ & $0.772 \pm 0.003$ \\
    \bottomrule
    \end{tabular*}
\end{table*}

Since we have good estimates for the orbital parameters and the brown dwarf to star mass ratio (from \citealt{Siverd2012}
RV observations), we can set strict priors on the Doppler beaming amplitudes for each passband. We calculate the beaming 
factors and beaming amplitude priors for each passband with \pytransit (namely using the \texttt{doppler\_beaming\_factor} 
and \texttt{doppler\_beaming\_amplitude} functions), using the BT-Settl \citep{Allard2013} model spectra to derive the 
beaming factors rather than a black body approximation. The beaming amplitude estimates are listed in 
Table~\ref{table:ev_and_db_priors}, and they are computed in the \ghlink{A2_doppler_beaming.ipynb }{Doppler beaming} notebook.

\subsubsection{Ellipsoidal variation}
\label{sec:analysis.priors.ev}

As with Doppler beaming, we can calculate theoretical EV amplitudes for all the passbands by combining our prior knowledge
about \nplanet and its host star with stellar spectrum models. We use the companion-star mass ratio estimate from \citet{Siverd2012}
and the semi-major axis estimate from \citet{Beatty2020}. We estimate the linear limb darkening coefficients
for each passband using \ldtk \citep{Parviainen2015b}, and estimate the gravity darkening coefficients for the \cheops passband
from the tables in \citet{Claret2020}, for the \tess passband from the tables in \citet{Claret2017}, and for the \pbh, \pbks, and 
\spitzer passbands from the tables in \citet{Claret2011}.\!\footnote{The EV amplitudes are calculated in the 
\ghlink{A1_ellipsoidal_variation.ipynb}{EV notebook} available from GitHub.}
As shown in Table~\ref{table:ev_and_db_priors}, the EV signal amplitude is 10-20 times larger than the DB signal amplitude, 
being roughly equal to the amplitude of the brown dwarf's phase curve signal in the \tess passband \citep{Beatty2020,VonEssen2021}.

The EV amplitude scales with the semi-major axis as $a_s^{-3}$, which makes the theoretical EV amplitudes sensitive
on our prior $a_s$ choice. This causes a potential problem because EV directly affects our eclipse depth measurements.
A biased prior $a_s$ estimate could also significantly bias the eclipse depths measured for passbands with only near-eclipse 
photometry (\cheops, \pbh, \pbks) if we were to base the EV amplitude priors on the theoretical amplitudes. Thus, 
it would be better to try to find a way to constrain the EV amplitudes in a way that is independent of prior mass 
ratio and semi-major axis estimates.

Fortunately, the \tess light curve covers the whole orbital phase with a precision that allows us to estimate the
EV amplitude in the \tess passband. The EV amplitude ratio between a passband $i$ and the \tess passband $t$
is simply
\begin{equation}
    \frac{A_\mathrm{EV,i}}{A_\mathrm{EV,t}} = \frac{\beta_\mathrm{i}}{\beta_\mathrm{t}},
\end{equation}
where the $\beta$ are the factors in Eq.~\eqref{eq:ev_beta} that depend on stellar gravity and limb darkening (which both 
are passband-dependent), but have no dependence on \nplanet's mass or orbit.

Thus, we can constrain the EV signal by constraining the EV amplitudes relative to \tess passband rather than constraining
the absolute EV amplitudes themselves. This approach is less likely to introduce biases to EV amplitudes, and so also less
likely to bias the secondary eclipse signal. In the final analysis, we set uninformative priors on the absolute 
EV amplitudes on all passbands, and set informative priors on the EV amplitude ratios between the \tess passband any every 
other passband
\begin{align*}
    \frac{A_\mathrm{EV,\,i}}{A_\mathrm{EV,\,t}}   &\sim N(\frac{\beta_\mathrm{i}}{\beta_\mathrm{t}}, \sigma),
\end{align*}
where the $\beta$ factors are calculated using the \ldtk and gravity darkening tabulations described earlier and 
$\sigma$ is the uncertainty in the $\beta$ ratio estimate. Further, instead of using the uncertainties listed in 
Table~\ref{table:ev_and_db_priors}, we set $\sigma$ to 0.01 to reduce our sensitivity to any biases in the 
$\beta$ factors (the chosen value is rather arbitrary but in general 3-10 times larger than the theoretical 
uncertainties).

\subsection{Noise and systematics model}
\label{sec:analysis.systematics}

In addition to the planetary signal and the EV and Doppler boosting signals from the star, the photometry contains
\emph{systematics} from astrophysical and instrumental sources, as well as from changes in the observing 
conditions. These systematics need to be taken into account during the phase curve modelling. 


We model the variability in the photometry not explained by the phase curve model as a Gaussian process
\citep{Rasmussen2006,Gibson2011a,Roberts2013b}. We use time as the only GP input for the \tess data, which 
allows us to use \celerite \citep{Foreman-Mackey2017} to evaluate the GP log likelihood for the whole 
unbinned \tess data set. The other data sets use more complex covariance functions with multiple input 
variables (e.g., x and y centroids, airmass and seeing), and for those we evaluate the GP log likelihood 
using \george \citep{Ambikasaran2014}.

Each GP adds a set of hyperparameters into the joint model. We use simple covariance kernels (Matern-3/2 and
squared exponential) which add only a log input scale parameter per input variable and a log output scale 
parameter per GP (detailed later separately for each data set). These GP hyperparameters are constrained
only loosely during the optimisation and MCMC sampling. We standardise the GP input variables (that 
is, we divide each input variable time series with its standard deviation and remove its mean) and constrain
the log input scale parameters using normal priors centred at one with a standard deviation of one.
The priors for the log output scale parameters are also normal and centred at log variance of the light curve 
with a standard deviation of 1.5.

We also fit the average white noise for each light curve. The white noise is parametrised using its log 
variance, and constrained with loose normal prior centred around a white noise estimate computed from the 
standard deviation of the flux point-to-point differences as
\begin{equation}
    \sigma = \mathrm{median}(\mathrm{sd}(\mathrm{diff}(\vec{f})))\; / \sqrt{2},
\end{equation}
where $\vec{f}$ is a vector containing the fluxes, sd stands for standard deviation, and $\mathrm{diff}$ stands 
for discrete difference.

The noise models are defined in the \texttt{\_init\_lnlikelihood} method of the \lpffn and 
\lpfed analysis classes, and the noise model parameter priors are set in the classes' 
\texttt{\_post\_initialisation} method. The posterior MCMC sampling marginalises over the whole
GP hyperparameter space allowed by the priors and the data to ensure we are not overfitting
the data.

\paragraph{\tess}
We model the variability in the \tess data as a Gaussian process with time as the only input variable 
using the \celerite package. The GP uses basic Matern-3/2 covariance
kernel that yields three additional parameters to the joint model, the log input and output scales
and the log white noise variance, and these parameters are constrained with relatively loose normal 
priors as described earlier.

\paragraph{LBT \pbh}
The systematics in the \pbh band light curve are modelled as a GP with eight input variables (including,
for example, the airmass and the target and comparison star locations). The covariance kernel is a product 
of eight squared exponential kernels where each kernel provides an independent input scale parameter into 
the joint model. This brings the total number of parameters from the \pbh band data set noise model to 
ten. 

\paragraph{CFHT \pbks}
The systematics in the \pbks band light curve are modelled as a GP with two input variables, the x and y
centroids. The covariance kernel is a product of two squared exponential kernels where each kernel provides 
an independent input scale parameter into the joint model, leading in total to four additional parameters.

\paragraph{\spitzer}
The systematics in the \spitzer data are also modelled using a GP with the x and y centroids as the
input variables. The \citet{Beatty2019} \spitzer phase curve observations were split into three 12~h 
long stares and the telescope was repointed between the stares. This repointing led to strong 
discontinuities in the raw photometry, so we decided to treat each 12~h stare as a separate
light curve. This gives us four light curves per \spitzer passband (one covering the \citealt{Beatty2013}
eclipse and three covering the phase curve). However, all the eight light curves use the same GP kernel 
and hyperparameters (we assume the centroid-related systematics in both \spitzer passbands have similar 
input and output scales), so the \spitzer GP model adds only four parameters to the joint model.

The \spitzer light curves consist of 10\;000 (eclipse) and 64\;000 (phase curve) flux measurements.
This is too much for a standard brute-force GP that requires inversion of the covariance matrix
and scales as $\mathcal{O}(n^3)$. Splitting the phase curve observations helps, but we still needed 
to bin the \spitzer observations to a two-minute cadence to make a GP noise model computationally 
feasible.

\paragraph{\cheops}
The \cheops data is modelled differently in the final analysis and in the initial \cheops-only analysis. 
In the \cheops analysis we use a linear model to model the systematics with 
basis vectors determined by the pipeline of \citet{Wilson2022} used as covariates. The number of basis 
vectors used per visit varies from 13 to 29, and the total number of basis vectors (and thus free parameters 
from the baseline model) is 156.

The noise model for the \cheops data assumes i.i.d. (independent and identically distributed) normally distributed noise (that is, white noise), and 
introduces only the logarithm of the average standard deviation of the noise distribution for each visit as 
an additional free parameter into the full model (that is, we assume the noise does not vary significantly
from observation to observation within a single light curve, and marginalise over the average observation
uncertainty). 

In the final analysis, we detrend the \cheops data using the posterior median baseline model from the \cheops
analysis and give each visit only a free normalisation term. We do this to simplify the final model, since 
using the full \cheops baseline model would add 156 free parameters into the final model. There is a danger
that this approach could lead to underestimated uncertainties in some of our quantities of interest, but 
a comparison of the \cheops and final analysis results (see Sect.~\ref{sec:analysis_scenarios}) shows
that any effects from not marginalising over the baseline model coefficients in the final analysis are
insignificant.

\section{Results and Discussion}
\label{sec:results_and_discussion}

\subsection{Results from the phase curve analysis}

\begin{figure*}
	\centering
	\includegraphics[width=\textwidth]{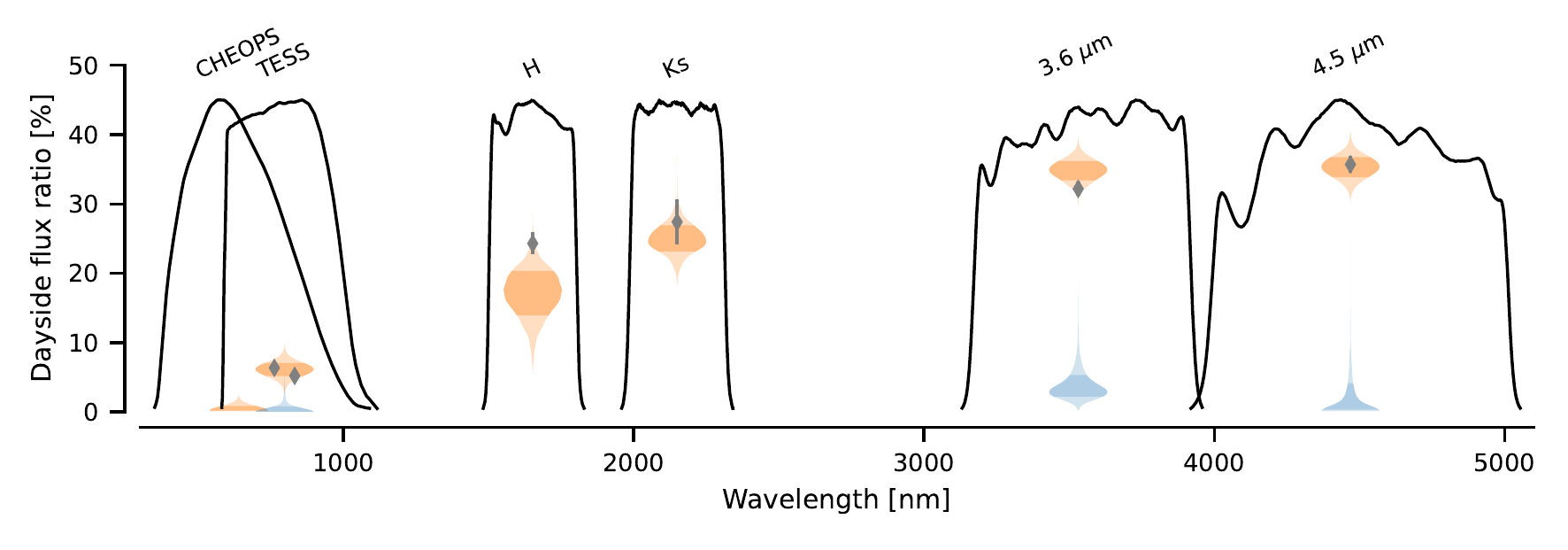}
	\caption{The dayside (orange) and nightside (blue) planet-star flux ratio posterior distributions for all passbands. 
	The central 68\% posterior interval is marked with a dark shade and the previous dayside flux ratio estimates
	are shown as diamonds with errorbars showing their reported uncertainties. For \tess, the \citet{Beatty2020}
	estimate is shown on the left and the \citet{VonEssen2021} estimate on the right.}
	\label{fig:day_and_night_fratio_posteriors}
\end{figure*}

\begin{figure}
	\centering
	\includegraphics[width=\columnwidth]{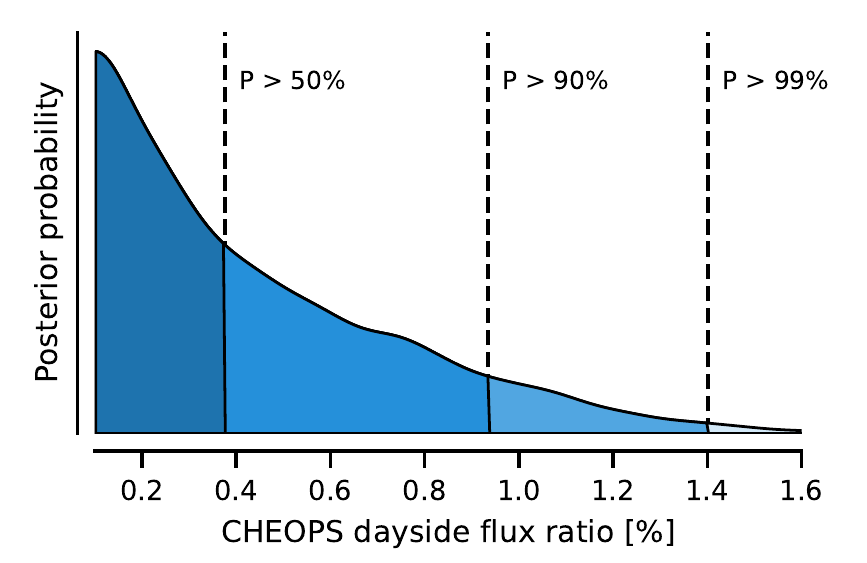}
	\caption{Dayside flux ratio posterior distribution for the \cheops passband. The dashed lines shows the posterior percentile limits.}
	\label{fig:cheops_fratio_posterior}
\end{figure}

\begin{figure*}
	\centering
	\includegraphics[width=\textwidth]{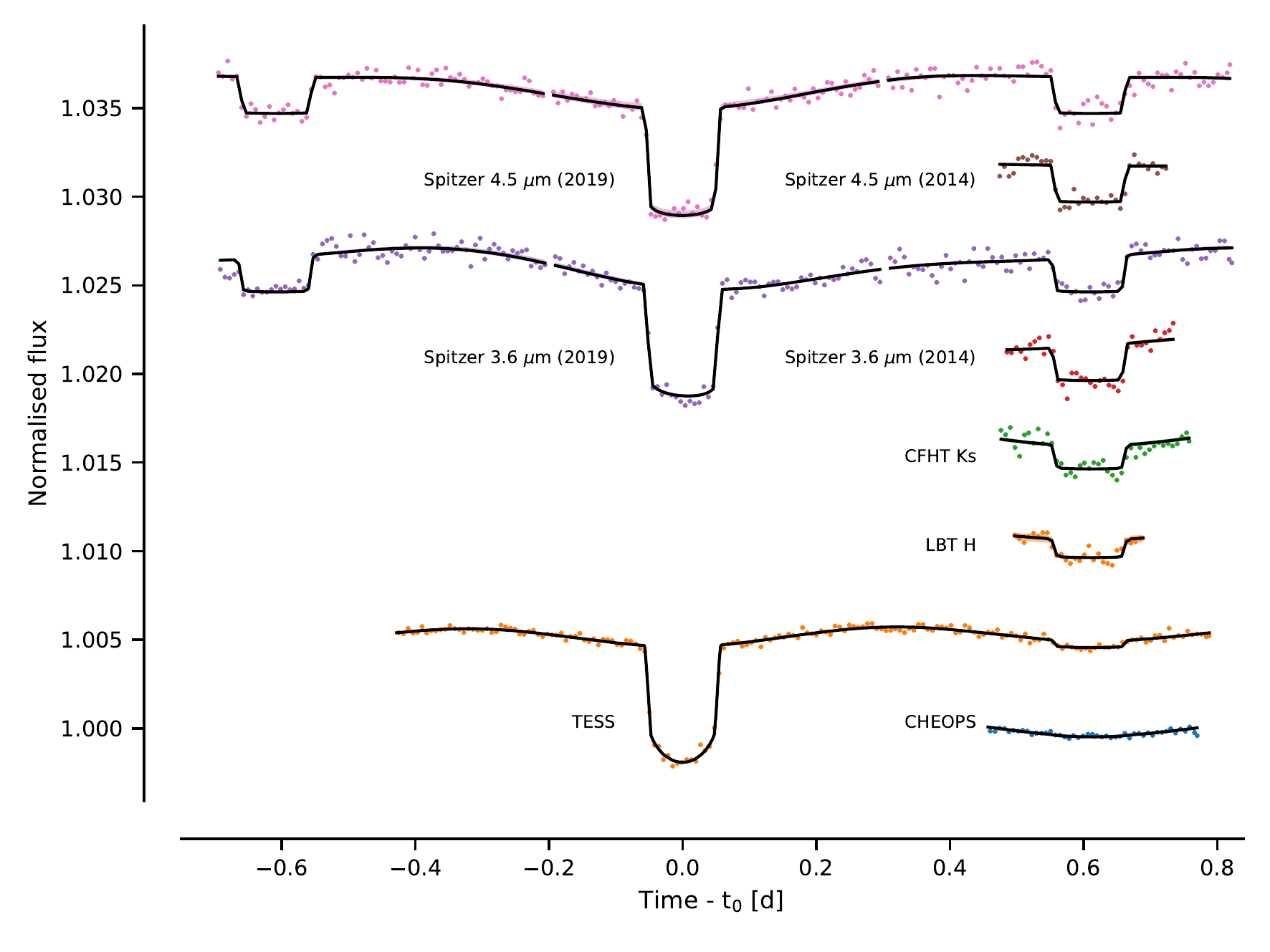}
	\caption{The observed light curves from \cheops, \tess, LBT, CFHT, and \spitzer (points) and the posterior 
	median phase curve model (black line). The systematics model has been removed from the observations, the 
	\tess and \cheops observations have been folded over the phase, and the data has been binned to 10 min for visualisation
	purposes. The figure is produced by the \ghlink{04_plots_final_light_curves.ipynb}{final light curve plot} notebook.}
	\label{fig:final_fit}
\end{figure*}

We show the dayside and nightside planet-star flux ratio posteriors from the final joint modelling for all passbands 
in Fig.~\ref{fig:day_and_night_fratio_posteriors}, give a more detailed view of the \cheops band dayside flux ratio 
posterior in Fig.~\ref{fig:cheops_fratio_posterior}, and show the observations and the model posteriors for all the passbands
in Fig.~\ref{fig:final_fit}. The final flux ratio estimates are reported in Table~\ref{table:flux_ratio_posteriors},
and the rest of the parameter estimates in Table~\ref{table:final_parameter_estimates}. 
The results from the external dataset and \cheops-only analysis scenarios are presented in Appendix~\ref{sec:analysis_scenarios}.

Except for the \cheops band, the dayside flux ratio posteriors are approximately normal with a well-defined non-zero mode, 
and for these posteriors we report their median values with uncertainties based on the 16th and 84th central percentiles in
Table~\ref{table:flux_ratio_posteriors}. The \cheops band flux ratio posterior has its mode at zero, and we report 
its posterior median and 99th percentile upper limits. 

The nightside flux ratio posteriors all have their modes at zero, and we report only the 99th posterior percentile upper
limits. The nightside flux ratio upper limit for the \tess band is 4.6\%, while for the \spitzer bands it is $\approx12\%$.

\begin{table*}
    \centering
    \caption{Final day- and nightside planet-star flux ratios and eclipse depths with their uncertainties and upper limits.
    }
    \label{table:flux_ratio_posteriors}
    \begin{tabular*}{\textwidth}{@{\extracolsep{\fill}} lccccc}
	\toprule\toprule
    Passband & Ns. flux ratio [\%] & Ns. \tbr [K] & Ds. flux ratio [\%] & Eclipse depth [ppm] & Ds. \tbr [K] \\
	\midrule     
    \cheops              &        &        &        $< 1.50$ &           $< 90$ & 2100 -- 2900 \\
    \tess                &  $< 4$ & < 3000 &  $ 6.1 \pm 0.9$ &  $  360 \pm  50$ & 3100 -- 3500 \\
    \pbh                 &        &        &  $17.3 \pm 3.2$ &  $ 1000 \pm 190$ & 2100 -- 3400 \\
    \pbks                &        &        &  $24.9 \pm 1.9$ &  $ 1460 \pm 110$ & 2900 -- 3400 \\
    \spitzer 3.6 $\mu$m  & $< 11$ & < 1900 &  $34.9 \pm 1.4$ &  $ 2040 \pm  80$ & 3000 -- 3400 \\
    \spitzer 4.5 $\mu$m  & $< 12$ & < 1800 &  $35.3 \pm 1.5$ &  $ 2070 \pm  80$ & 2800 -- 3200 \\
    \bottomrule
    \end{tabular*}
    \tablefoot{The nightside estimates give an upper limit corresponding to the 99th posterior percentiles. The dayside
    estimates correspond to posterior medians with uncertainties based on the 16th and 84th posterior percentiles (except
    for \cheops, for which we only give the upper limit). The reported dayside brightness temperature values are lower and 
    upper limits corresponding to 2.5th and 97.5th posterior percentiles.}
\end{table*}

\begin{table*}
	\centering
	\caption{Final estimates for the geometric and orbital parameters of \nplanet and its host star.}
	\label{table:final_parameter_estimates}  
	\begin{tabular*}{\textwidth}{@{\extracolsep{\fill}} llll}        
		\toprule\toprule
		\multicolumn{4}{l}{\emph{Ephemeris}} \\
		\midrule
		Transit epoch & $T_0$ & [BJD] & $2455914.1624 \pm 4 \times 10^{-4}$\\
		Orbital period & $P$ & [days] &  $1.2174942 \pm 2\times 10^{-7}$ \\
		Transit duration & $T_{14}$ & [h] & $2.74 \pm 0.01$ \\
		\\
		\multicolumn{4}{l}{\emph{Relative properties}} \\
		\midrule
        Area ratio & $k^2$ &$[A_\star]$ & $0.00585 \pm 7 \times 10^{-5}$ \\
        Radius ratio &$k$ & $[R_\star]$ & $ 0.0765 \pm 5 \times 10^{-4}$ \\
		Semi-major axis &$a$ & $[R_\star]$ & $3.53 \pm 0.09$ \\
		Impact parameter &$b$ & $[R_\star]$ & $0.35\;(-0.08)\;(+0.06)$ \\
		\\
		\multicolumn{4}{l}{\emph{Absolute properties}} \\
		\midrule 
		\nplanet radius &$R_{\mathrm{c}}$ & [\rjup]  &  $ 1.138 \pm 0.010 $ \\
		\nplanet surface gravity & $\log g_\mathrm{c}$ & [cgs] & $4.73 \pm 0.02$ \\ 
		\nplanet eq. temperature$^a$ & $T_{\mathrm{eq}}$ &[K] & $ 2550 \pm 170 $ \\
		Semi-major axis &$a$ &[AU] & $ 0.0251 \pm 0.0006 $\\
		Orbital inclination & $i$ &[deg] & $84.3 \; (-1.1) \; (+1.4) $ \\
		Stellar density & $\rho_\star$ & $[\gcm]$ & $0.56 \pm 0.04 $\\
		\tess emission offset & $O_\mathrm{T}$& [deg] & $3 \pm 9$ \\
		\spitzer \pbsa emission offset & $O_\mathrm{S1}$ & [deg] & $-29 \pm 5$ \\
		\spitzer \pbsb emission offset & $O_\mathrm{S2}$ & [deg] & $4 \pm 6$ \\
		\bottomrule       
	\end{tabular*}
	\tablefoot{The estimates correspond to the posterior median with the uncertainties based on the 16th and 84th posterior percentiles.
		\tablefoottext{a}{The equilibrium temperature is calculated using the stellar \teff{} estimate, 
		scaled semi-major axis distribution, heat redistribution factor distributed uniformly between 0.25 and 0.5, and
		albedo distributed uniformly between 0 and 0.4.}}
\end{table*}

The dayside flux ratios for the \tess and \spitzer passbands (Fig.~\ref{fig:day_and_night_fratio_posteriors}) agree fairly well 
with the previous studies by \citet{Beatty2020}, \citet{VonEssen2021}, and \citet{Beatty2019}. The flux ratios for the 
\pbh and \pbks bands are lower than what reported by \citet{Beatty2017} and \citet{Croll2015}, but this is expected since 
neither of the original studies included an ellipsoidal variation signal into their models. 

The dayside flux ratio for the 3.6~$\mu$m \spitzer band is higher than the estimate by \citet{Beatty2019}. 
The \spitzer light curves use relatively flexible GP systematics models and all the \spitzer light curves are well-fit visually 
(Fig.~\ref{fig:final_fit}). 
Even then, we cannot rule out poorly modelled systematics as the source of the anomalously high 3.6~$\mu$m flux ratio 
since the pre-eclipse baseline is relatively short for the full-phase \spitzer observations.

\begin{figure}
	\centering
	\includegraphics[width=\columnwidth]{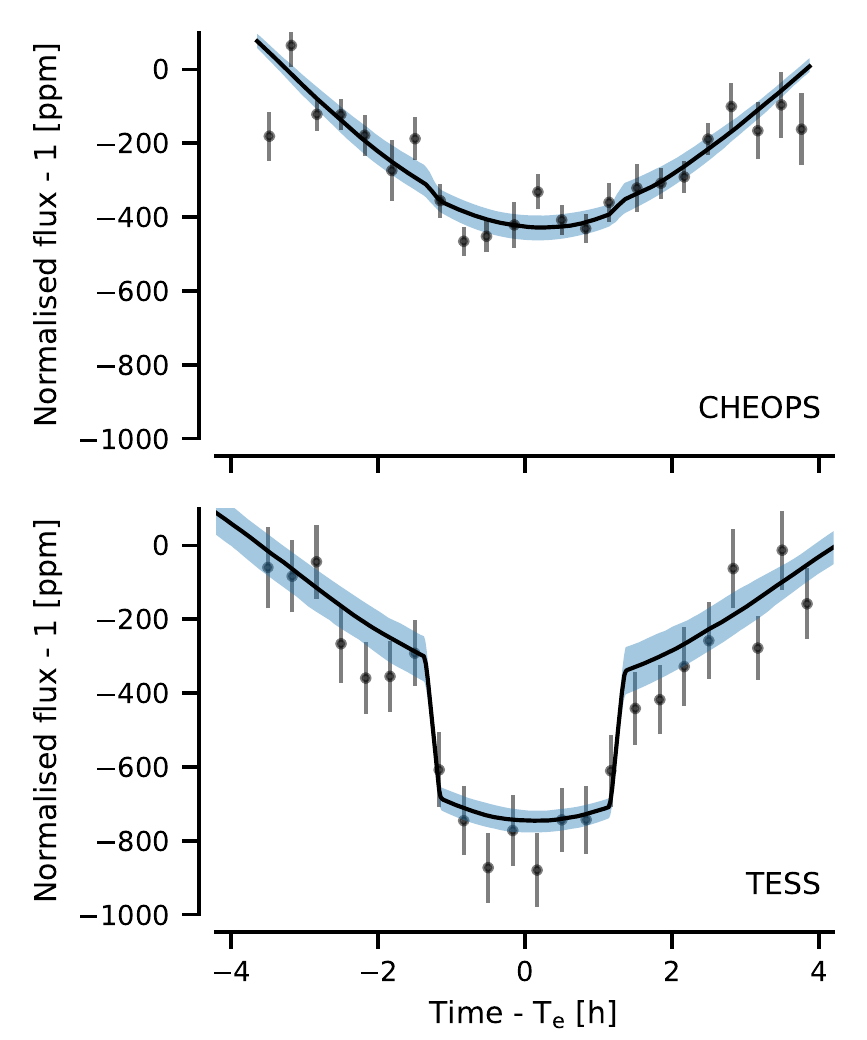}
	\caption{The \cheops eclipse observations binned to a 20-minute cadence with the phase curve model (panel a) and 
	\tess observations with a similar binning (panel b). The median posterior phase curve model is shown 
	as a black line and its 68\% central posterior interval is shown as a shaded blue region.}
	\label{fig:cheops_light_curve}
\end{figure}

Curiously, the \cheops and \tess bands show a strong discrepancy in their dayside flux ratios. Panel a) of
Fig.~\ref{fig:cheops_light_curve} shows the \cheops observations with the median posterior model and panel 
b) shows the same for \tess. The combined \cheops observations have a noise estimate of $\approx50$~ppm over 
a 20-min bin, half of the \tess noise of $\approx97$~ppm, but we do not detect a significant eclipse signal
in the \cheops band. Instead, we are only able to set an upper limit of 1.4\% for the \cheops dayside flux ratio, 
while in the \tess band we detect a clear eclipse corresponding to a flux ratio of $6\pm1\%$. 

\subsection{Dayside brightness temperatures}
\label{sec:discussion.brightness_temperatures}

\begin{figure}
	\centering
	\includegraphics[width=\columnwidth]{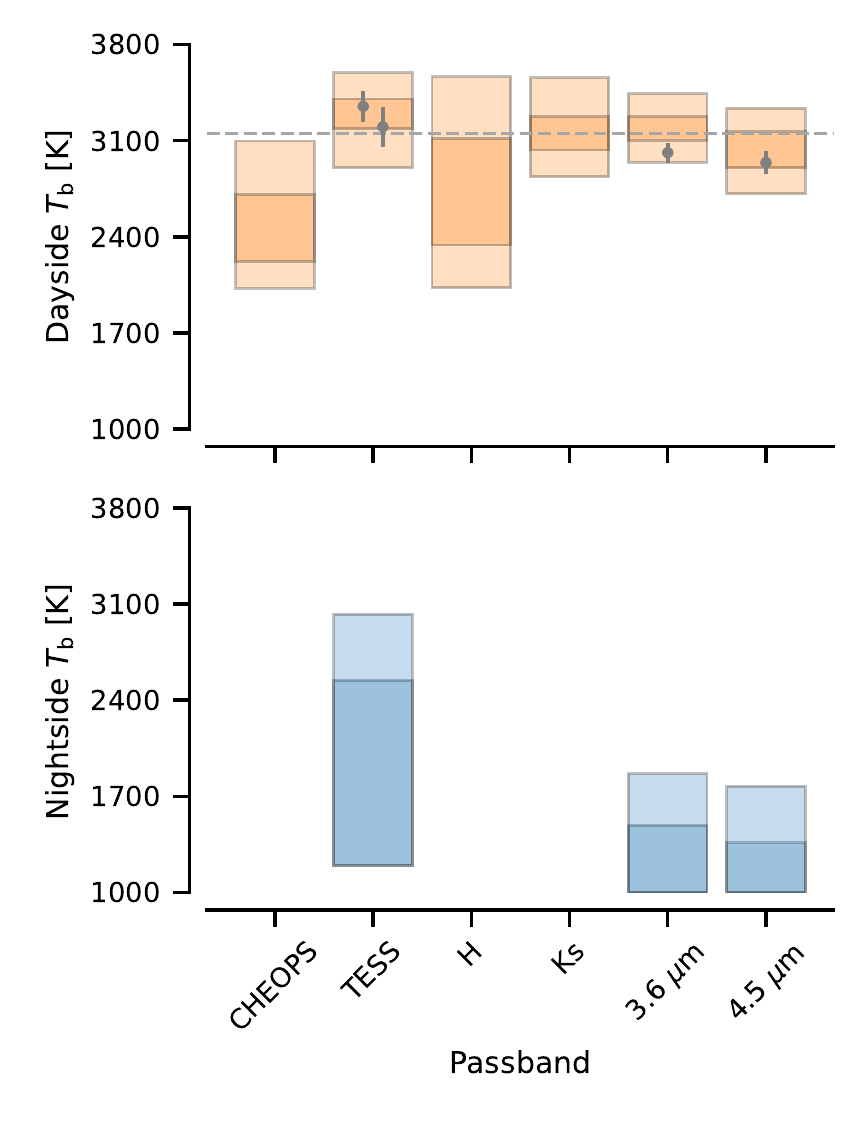}
	\caption{Day- and nightside brightness temperatures estimated from the flux ratios using the BT-Settl stellar model 
	spectra by \citet{Allard2013}. The shading marks the 68\% and 99.7\% central posterior intervals, from the
	darkest to the lightest shade, and the dashed horizontal line shows the best-fitting constant brightness temperature.
	The previous dayside \tbr estimates for the \tess and \spitzer passbands by \citet{Beatty2019}, \citet{Beatty2020}, 
	and \citet{VonEssen2021} are shown as points with errorbars.}
	\label{fig:brigthness_temperatures}
\end{figure}

We estimate the day- and nightside brightness temperatures, \tbr, separately for each passband using the 
flux ratio estimates from the final joint modelling and  BT-Settl stellar spectrum models by \citet{Allard2013}. 
We calculate the \tbr estimates as averages over three approaches: a) physical-physical, where both the brown dwarf 
and its host star spectra are modelled with BT-Settl models; b) blackbody-physical, where  \nplanet is modelled 
as a black body and the star uses a BT-Settl model; and c) blackbody-blackbody, where both \nplanet and the star 
are modelled as black bodies. 

Figure~\ref{fig:brigthness_temperatures} shows the approach-averaged brightness temperatures together with the
best-fitting constant dayside temperature (the \tbr estimates are also included in Table~\ref{table:flux_ratio_posteriors},
and the per-approach estimates are shown in Appendix~\ref{sec:app.brightness_temperatures}). The dayside temperatures mostly agree
with each other within the uncertainties. In agreement with \citet{Beatty2020}, the \tess passband has a somewhat 
higher \tbr than the \spitzer bands, but the disagreement is less significant for us (for example, the \tess and 
\spitzer \pbsa band 68\% central posterior intervals overlap). Our \tbr estimate for the \tess passband agrees
well with the \citet{Beatty2020} result, but our \spitzer \tbr estimates are higher than what reported in 
\citet{Beatty2019}.

As expected from the differences in the dayside flux ratios, the dayside \tbr estimate for \cheops is significantly 
lower than for \tess.

\subsection{Nightside brightness temperatures}

We can measure the nightside brightness temperatures only for the passbands where we have observations over the
full orbital phase (\tess and \spitzer), and the estimates are rather poorly constrained. Both \spitzer passbands 
support cool night sides, with 99th \tbr posterior percentiles $\approx 1850~K$, while in the \tess passband we
can only set a 3000~K upper limit. These results support a significant temperature difference between the night-
and dayside, and agree with the previous \spitzer and \tess analyses by \citet{Beatty2020} and \citet{VonEssen2021}

\subsection{Atmospheric modelling}
\label{sec:discussion.atmospheric_modelling}

For the atmospheric modelling, we follow the same basic approach as done in \citet{Lendl2020}. In particular, we employ the self-consistent atmosphere model \texttt{HELIOS} \citep{Malik2017, Malik2019} to generate a grid of possible atmospheres for \nplanet. As discussed in \citet{Lendl2020}, in the context of 1D models, the flux of the host star impinging upon the top of the atmosphere (TOA) of \nplanet is \citep{Cowan2011} 
\begin{equation}
  F_{\mathrm{TOA}} = F_* \left( \frac{R_*}{a} \right)^2 \left(1 - A_\mathrm{B} \right) \left(\frac{2}{3} - \frac{5\epsilon}{12} \right) \ ,
\end{equation}
where $F_*$ is the stellar photometric flux, $R_*$ the radius of the host star \nstar, $a$ the orbital distance of \nplanet, $A_\mathrm{B}$ its Bond albedo, and $\epsilon$ the heat redistribution efficiency ($0\leq \epsilon \leq 1$).
In terms of the atmosphere model \texttt{HELIOS}, the term
\begin{equation}
  E_* = \left(1 - A_\mathrm{B} \right) \left(\frac{2}{3} - \frac{5\epsilon}{12} \right)
\end{equation}
is fully degenerate with respect to $A_\mathrm{B}$ and $\epsilon$, such that we only vary $E_*$ rather than $A_\mathrm{B}$ and $\epsilon$ independently. 

Since \nplanet is a brown dwarf, its emission might also have a considerable contribution from an internal energy flux, described by an internal temperature $T_\mathrm{int}$. In case of an isolated object without exterior insolation, $T_\mathrm{int}$ would be the object's effective temperature. For the construction of the model grid, we therefore use $T_\mathrm{int}$ as an additional free parameter. Given the lack of any spectrally-resolved data for \nplanet and the fact that the host star's [Fe/H] value is consistent with 0 (see Table \ref{table:star}), we use solar element abundances for the model grid, for simplicity. The general modelling parameters for \nplanet (surface gravity, stellar radius, orbital distance) are taken from Tables \ref{table:star} \& \ref{table:final_parameter_estimates}. 

Using \texttt{HELIOS}, we generate a model grid of self-consistent forward atmosphere models for \nplanet in terms of the free parameters $T_\mathrm{int}$ and $E_*$. In total we compute 165 different atmospheric model structures for internal temperatures between 0 K and 3000 K and $E_*$ values from 0.667 to 0.06. The former value is the highest possible one for $E_*$, with a Bond albedo and heat redistribution $\epsilon$ of 0. The latter value represents a case with a high Bond albedo and/or very efficient heat redistribution.

Besides the atmospheric structure, we also obtain the brown dwarf's emission spectrum for each model in a postprocess procedure. For this postprocess, we exclude any scattering contribution, such that the obtained emission spectra describe only the atmosphere's thermal emission. These spectra are then integrated over the passbands shown in Table \ref{table:flux_ratio_posteriors}
\begin{equation}
  F_\mathrm{pb} = \int \mathcal F F_\lambda \, \mathrm d \lambda 
\end{equation}
to calculate the flux $F_\mathrm{pb}$ inside each passband. Here, $\mathcal F$ is the corresponding filter transmission function and $F_\lambda$ the calculated emission spectrum of KELT-1b. 

Since these $F_\mathrm{pb}$ are a smooth function of $T_\mathrm{int}$ and $E_*$, we parameterise them as a two-dimensional, fifth-order polynomial in each passband. This allows us to evaluate the passband-integrated fluxes $F_\mathrm{pb}(T_\mathrm{int}, E_*)$ as a continuous function of $T_\mathrm{int}$ and $E_*$ directly, rather than having to perform an interpolation in two dimensions. 

These passband-integrated fluxes are then added to a forward model in the Bayesian retrieval framework \texttt{Helios-r2} \citep{Kitzmann2020} to constrain the geometric albedos in the \tess and \cheops passbands ($A_\mathrm{g,T}$ and $A_\mathrm{g,C}$). This forward model calculates the secondary eclipse depths $\mathrm d F_\mathrm{se}$
\begin{equation}
  \mathrm d F_\mathrm{se} = A_g \left(\frac{R_\mathrm{p}}{a} \right)^2 + \frac{F_\mathrm{pb}(T_\mathrm{int}, E_*)}{\int \mathcal F F_* \, \mathrm d \lambda} \left(\frac{R_\mathrm{p}}{R_*} \right)^2
\end{equation}
as a function of the free parameters $A_g$, $R_\mathrm{p}/a$, $R_\mathrm{p}/R_*$, $T_\mathrm{int}$, and $E_*$ in each photometric passband. The stellar photospheric spectrum $F_*$ is taken from the grid of PHOENIX stellar atmosphere models \citep{Husser2013}, based on the stellar parameters given in Table \ref{table:star}.

Within the retrieval forward model we assume that only the \cheops and \tess passbands have contributions by a geometric albedo, whereas the other ones are dominated by thermal emission rather than reflected light. This is based on the fact that the stellar spectrum peaks at lower wavelengths, where the \cheops and \tess passbands are located. The \tess band
receives 64\% of the stellar flux received by the \cheops band, and the \pbh, \pbks, \pbsa, and \pbsb bands receive 7\%, 3\%, 1\%, and 0.5\% of the flux received by the \cheops band, 
respectively. The four photometric measurements (\pbh, \pbks, \spitzer) in the infrared are dominated by the brown dwarf's own thermal emission rather than by scattering 
of its host star's incident flux.

\begin{figure}
	\centering
	\includegraphics[width=\columnwidth]{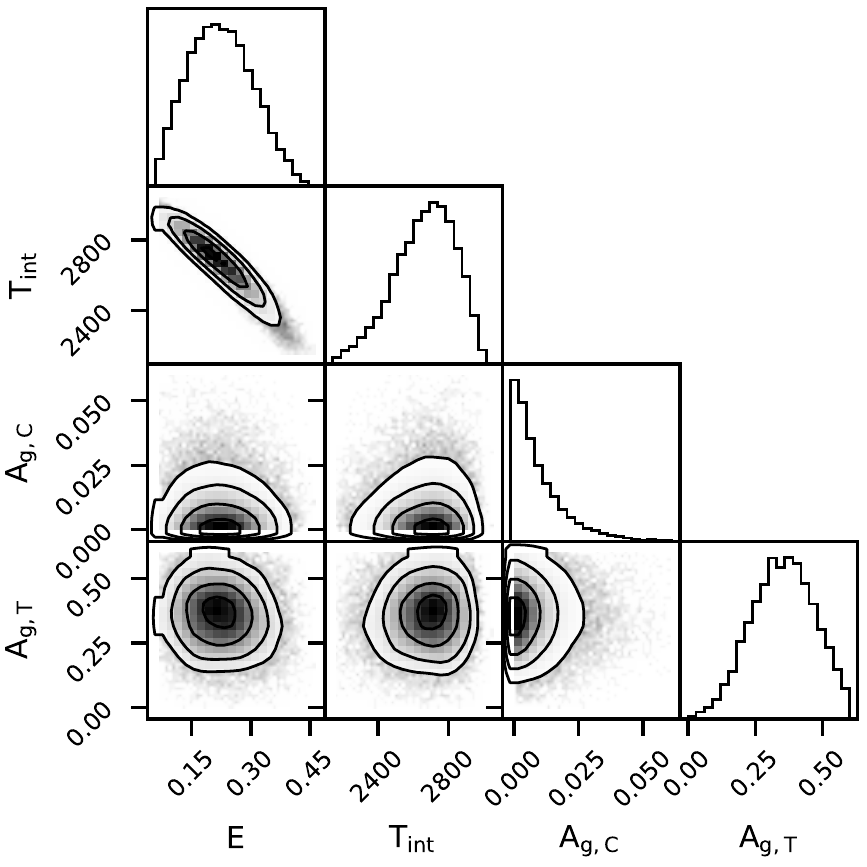}
	\caption{Posterior distributions for the retrieval calculations using \texttt{Helios-r2}. We show only the four most important
	parameters. The rest are strongly constrained by priors and do not show significant correlations between other parameters.}
	\label{fig:retrieval_posteriors}
\end{figure}

\begin{figure*}
	\centering
	\includegraphics[width=\textwidth]{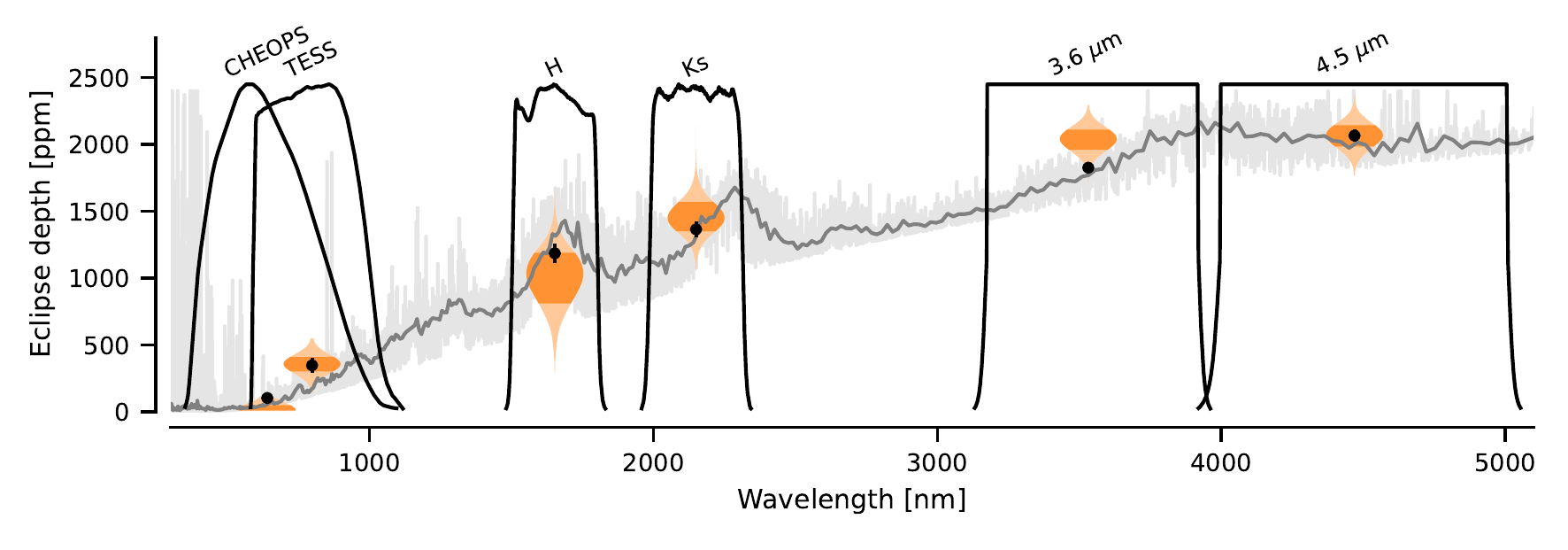}
	\caption{Posterior model spectrum for the retrieval calculations using \texttt{Helios-r2}. The light gray line shows the high resolution
	version of the posterior thermal emission model median, and the darker gray line shows the same spectrum in medium resolution. 
	The black points show the retrieval eclipse depths with their corresponding 1$\sigma$ intervals, and the orange shaded areas show 
	the eclipse depth posteriors estimated from the photometry as given in Table~\ref{table:flux_ratio_posteriors}. The black labelled lines
	show the passbands, where the \spitzer passbands are simplified for visualisation.}
	\label{fig:retrieval_spectrum}
\end{figure*}

The posterior distributions of our retrieval calculations are shown in Fig. \ref{fig:retrieval_posteriors} while the posterior photometric points are depicted in Fig.~\ref{fig:retrieval_spectrum}. The posterior distributions show a strong correlation between the parameters $E_*$ and $T_\mathrm{int}$. 
When the retrieval is performed only on the TESS and Spitzer data points, bimodal solutions for these two quantities are obtained (see Appendix \ref{sec:additional_retrievals}).  This bimodality is broken primarily by the \pbh and \pbks data points, while the \cheops data point further narrows the posterior distribution on $E_*$.
The retrieved value of $T_\mathrm{int} = 2670_{-150}^{+190}$~K implies that the brown dwarf has a very strong contribution from an internal heat flux to the total energy budget of the atmosphere and corroborates the findings of \cite{Beatty2020}. As already noted by \cite{Beatty2019}, an isolated brown dwarf with the same age and mass would have an internal temperature of about 850 K. The relatively low value of $E_* = 0.22 \pm 0.08$, on the other hand, suggests that the heat recirculation $\epsilon$ must very high, or that the atmosphere has a large Bond albedo $A_\mathrm{B}$. 

For the \tess passband, we obtain a geometric albedo of about $0.36^{+0.12}_{-0.13}$, while in the \cheops passband the results are consistent with a geometric albedo of 0.

The resulting photometry posteriors in comparison to the measured data are shown in the lower panel of Fig.~\ref{fig:retrieval_posteriors}. The results imply that atmosphere forward models can explain most of the measured data points while especially the \spitzer 3.6 $\mu$m band shows larger deviations. This mismatch in the \spitzer 3.6 $\mu$m band might point to non-equilibrium chemistry effects. The predicted \cheops photometry secondary eclipse depth also deviates from the measured one. In fact, even without including an additional geometric albedo contribution, the measured eclipse depth cannot be described by the expected thermal emission of the planet.

\begin{figure}
	\centering
	\includegraphics[width=\columnwidth]{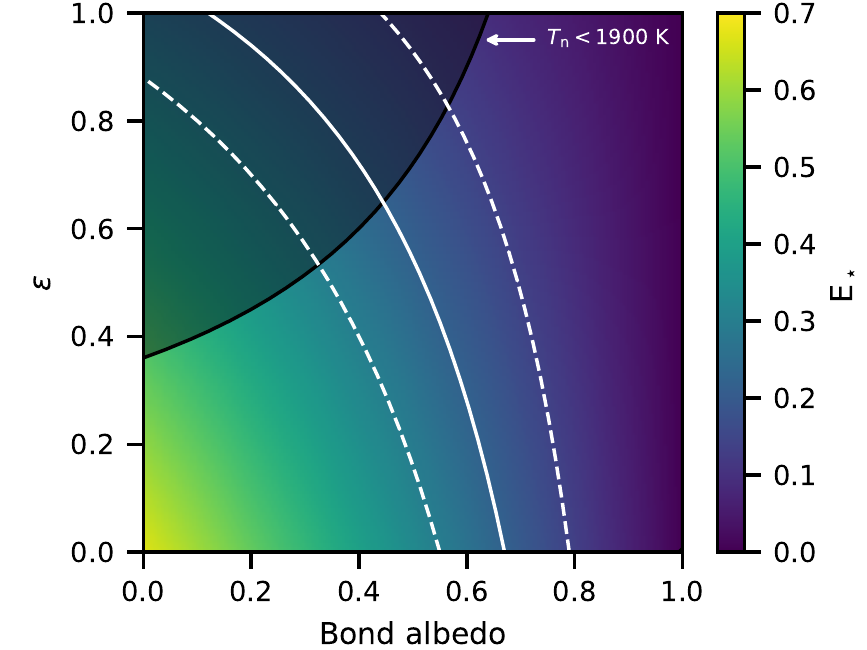}
	\includegraphics[width=\columnwidth]{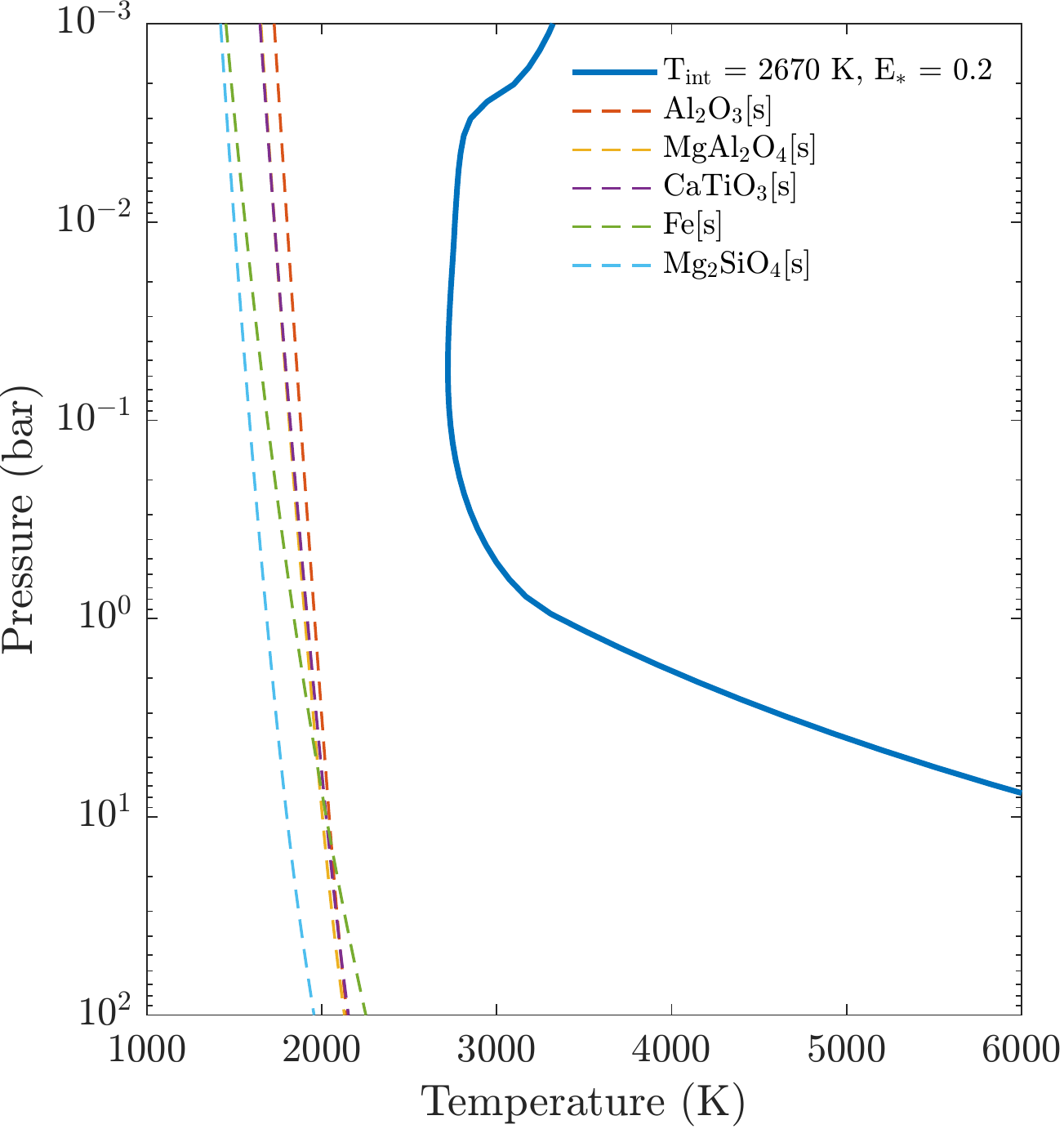}
	\caption{The upper panel depicts the values of $E_*$ as function of the heat recirculation parameter $\epsilon$ and Bond albedo $A_\mathrm{B}$. The solid white line represents the retrieved median value of $E_*$, the dashed lines are its corresponding 1$\sigma$ interval, and the shaded region corresponds to the parameter space excluded by the upper nightside \tbr limit of 1900~K. The lower panel shows the temperature-pressure profile of a \texttt{HELIOS} model (solid line), computed for the median values of $E_*$ and $T_\mathrm{int}$. Besides the temperature-pressure profile, the lower panel also includes  a set of condensate stability curves (dashed lines) for a selected set of species.}
	\label{fig:retrieval_single_models}
\end{figure}

The upper panel of Fig. \ref{fig:retrieval_single_models} depicts the heat recirculation efficiencies $\epsilon$ and Bond albedos $A_\mathrm{B}$ that are consistent with the retrieved value of $E_* = 0.22 \pm 0.08$ and the upper nightside \tbr limit of 1900~K. If the heat redistribution is very efficient ($\epsilon \approx 1$), then the Bond albedo has to be close to 0. On the other hand, if only very little heat gets transported to the nightside ($\epsilon \approx 0$), then the Bond albedo has to be in the range of 0.7, implying that potentially clouds scatter almost 70\% of the incident stellar radiation back to space. In principle, given constraints on $T_{\rm int}$, the dayside temperature and the nightside temperature, the degeneracy between $A_{\rm B}$ and $\epsilon$ could be broken.

Additionally, we calculate a separate \texttt{HELIOS} model for the above stated retrieved median values of $E_*$ and $T_\mathrm{int}$. The resulting temperature-pressure profile is shown in the bottom panel of Fig. \ref{fig:retrieval_single_models}. Due to the very high $T_\mathrm{int}$, the temperature increases very strongly in the lower atmosphere, while being close to isothermal around 0.1 bar, and showing a temperature inversion in the upper atmosphere, originating from the absorption of stellar light. The plot also depicts a set of stability curves for selected high-temperature condensates. These stability curves in combination with the atmosphere's temperature profile imply that most likely condensates do not form on the dayside of KELT-1b. This is not surprising given its high equilibrium temperature.

The derived geometric albedo of $0.36^{+0.12}_{-0.13}$ in the \tess passband is challenging to explain, since even the most refractory mineral (corundum) does not condense at the derived dayside temperatures. However, it agrees with the scenario where silicate clouds form on the nightside of the brown dwarf and are then transported to the dayside by winds, as well as with geometric albedos for silicate cloud-covered planets of $\sim0.4$ predicted by \citet{Sudarsky2000a}. It is nevertheless difficult to reconcile this high value of the \tess geometric albedo with a \cheops geometric albedo of essentially zero, because the two passbands have considerable overlap.

\subsection{Chromatic albedo variability as a possible source of discrepancy between \cheops and \tess}
\label{sect:discussion.albedo}

The albedo spectrum of a planet (or, as in our case, a brown dwarf) can feature clear distinct regions of low and high reflectivity
in wavelength space \citep{Sudarsky2000a,Burrows2008}. In theory, a low blue-optical albedo combined with a high NIR albedo could be 
explained with silicate clouds and a strong optical absorber above the silicate cloud deck \citep[such as gaseous TiO/VO or 
S$_2$/HS,][]{Schwartz2015}, and there is a possibility that the discrepancy between the \cheops and \tess dayside flux ratios
could be explained by a high-contrast feature in \nplanet's albedo spectrum. 

We modelled whether the \cheops-\tess discrepancy could be explained by chromatic variations in 
\nplanet's albedo using a simple two-level albedo spectrum model parametrised by the geometric albedo in blue, \agb, 
geometric albedo in red, \agr, and step location, $\lambda_0$, dictating where the blue albedo changes to red. The modelling 
is detailed in Appendix~\ref{sec:app.chromatic_albedo} and carried out in the \ghlink{A3_chromatic_albedo_variation_test.ipynb}{chromatic 
albedo variation test} notebook.

As a conclusion, the discrepancy is challenging to explain with a simple two-level albedo spectrum model due to the large overlap 
between the \cheops and \tess bands.
The largest \tess to \cheops reflected light contrast, $C = (F_\mathrm{T}-F_\mathrm{C})/(F_\mathrm{T}+F_\mathrm{C})$,  
of $0.41$ is obtained for a step model with $\ag = 0.01$ for $\lambda < 860$~nm and $\ag = 0.4$ for $\lambda \geq 860$~nm, 
as illustrated in Fig.~\ref{fig:c_vs_t_albedo_1}. A geometric albedo of $0.01$ is not unheard of for a highly irradiated
body, but is similar to what has been estimated for TrES-2b \citep{Kipping2011b}, while the geometric albedo of $0.4$ 
corresponds to the silicate cloud models by \citet{Sudarsky2000a}, and is in line with the previous \nplanet study by \citet{Beatty2020}. 
However, the maximum contrast of 0.4 is at the lower tail of the measured $C$ posterior distribution that has its
mode at 0.9 (Fig.~\ref{fig:c_vs_t_albedo_2}). Even a blue \ag of 0.001 with $\lambda_0 = 940$~nm would lead to $C$ of 
only 0.6, so it is unlikely that a physically plausible albedo spectrum would be able to reproduce the observed 
\cheops-\tess discrepancy.

\subsection{Temporal variability as a possible source of discrepancy between \cheops and \tess}
\label{sect:discussion.variability}

Since the upper limits on the nightside temperature are low enough that clouds could conceivably form, it is plausible 
that nightside clouds could be transported to the dayside \citep{Beatty2020}, where they could survive transiently. This 
raises the possibility of dayside cloud cover being transient. Variability in cloud cover could be one possible explanation 
for the dayside flux ratio discrepancy between the
\cheops and \tess passbands. \citet{VonEssen2021} reported possible temporal variation in the eclipse depths in the
\tess data, but concluded that the variations were most likely associated with stellar variability. We likewise carried 
out per-eclipse analyses for the \tess and \cheops data, but also need to conclude that the per-eclipse precision
is not high enough to study whether the eclipse depths show systematic variations that would securely not be caused
by stellar variability or systematics.

The spatial distribution of transient dayside clouds transported from the brown dwarf's nightside can be expected to be 
strongly nonuniform, which would lead to asymmetric reflection phase curve as with Kepler-7b \citep{Demory2013}. While our analyses
simplify the \nplanet's flux contribution by using only the emission component, the emission component has a loosely
constrained phase offset that allows it to catch any existing phase curve asymmetries. 

The emission offset (see Table \ref{table:final_parameter_estimates}) agrees with zero for the \tess and \spitzer \pbsb bands, but is 
significantly non-zero for the \spitzer \pbsa band. Of these three, only the \tess band phase curve is expected to have a significant 
contribution from reflection, and the lack of measurable phase curve asymmetry speaks against the cloud transport hypothesis. Then again,
were the dayside cloud coverage to be variable, the variability could blur the asymmetry when averaged over the \tess observations of 25 days.

The \tess and \spitzer phase offsets disagree with the ones estimated by \citet{Beatty2019} and \citet{Beatty2020}. They observe
slight eastward shifts for the \tess and \spitzer \pbsb bands and a strong eastward shift for the \spitzer \pbsa band, while our
shifts are westwards.

The phase curve asymmetry in \spitzer \pbsa band is curious. The \spitzer phase curve observations were carried out eight days apart of 
each other, so the differences could be explained by temporal variability in emission and reflection. This could also explain the higher-than
expected dayside eclipse depth for the \pbsa band (Fig.~\ref{fig:retrieval_spectrum}). Our eclipse depth estimate differs from the 
result from \citet{Beatty2019} (whose estimate agrees with the theoretical models) and we extracted the \spitzer light curves
using our own pipeline, so we cannot completely rule out untreated systematics. However, the time-scale of the variability (smooth
over \nplanet phase) combined with our use of a flexible GP to model the \spitzer systematics lead us to believe the differences
in the two \spitzer phase curves arise from true variability in the \nplanet's phase curve.

\subsection{Comparison between emission spectrum models}

\begin{figure*}
	\centering
	\includegraphics[width=\textwidth]{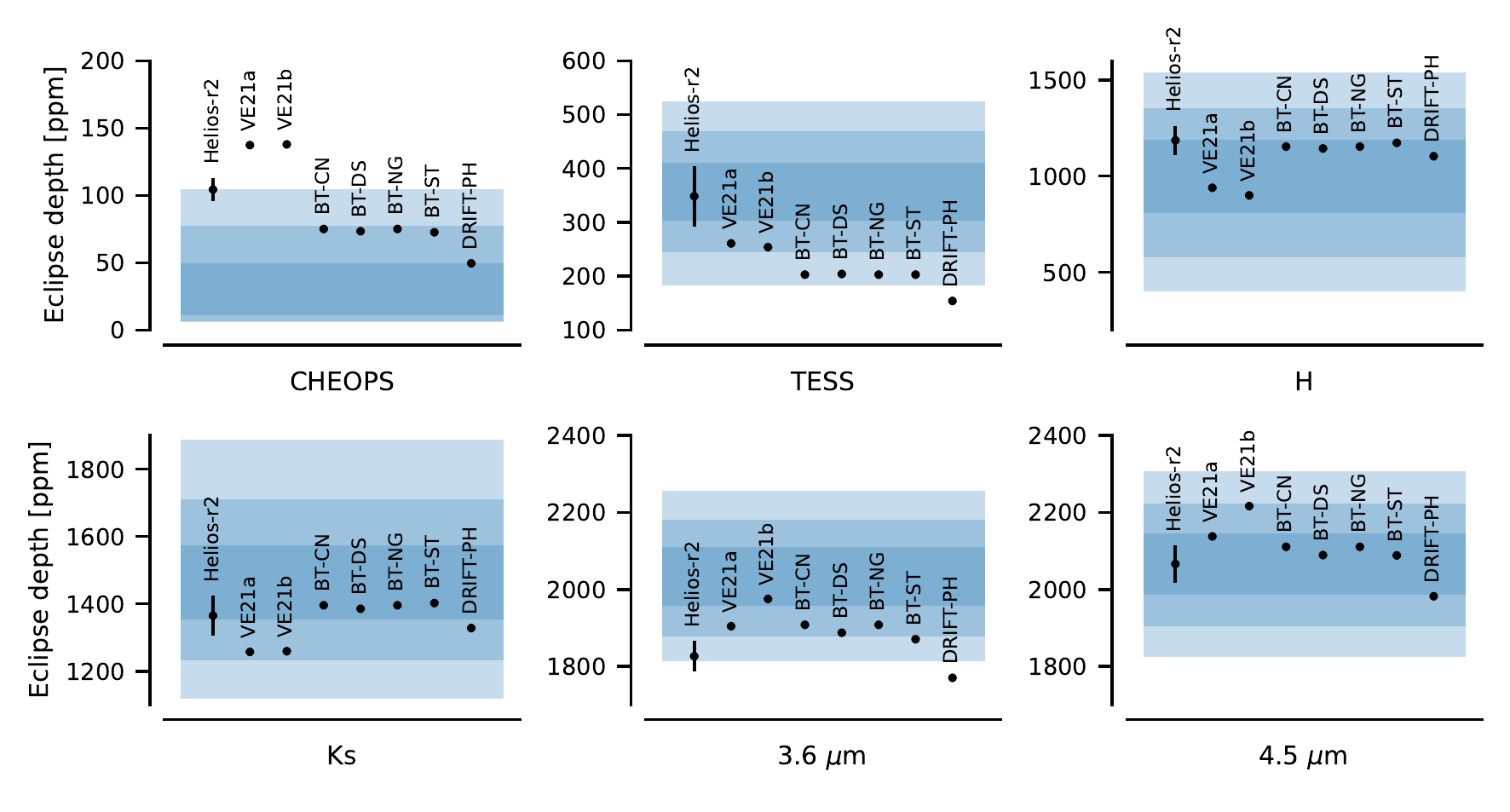}
	\caption{Comparison between different emission spectrum models. The shaded areas show the 68\%, 95\%, and 99.7\%
	central posterior intervals for the eclipse depths as estimated from the joint modelling, and the black points show
	the passband-integrated emission spectrum models. Helios-r2 corresponds to the model presented here, VE21a and VE21b to the models in \citet{VonEssen2021},
	BT-CN, BT-DS, BT-NG, and BT-ST to the best-fitting BT-Cond, BT-Dusty, BT-NextGen, and BT-Settl models by \citet{Allard2013}, respectively,
	and DRIFT-PH to the best-fitting Drift-PHOENIX model \citep{Witte2011}.}
	\label{fig:spectrum_comparison}
\end{figure*}

We compare our retrieved emission spectrum with the spectra by \citet{VonEssen2021}, the \texttt{BT} model spectra
\citep{Allard2013}\footnote{Downloaded from F.~Allard's \href{https://phoenix.ens-lyon.fr/Grids/}{Star, Brown Dwarf, and Planet simulator}.} 
and \texttt{DRIFT-PHOENIX} spectra \citep{Witte2011}\footnote{Obtained from the 
\href{http://svo2.cab.inta-csic.es/theory/newov2/index.php}{SVO Theoretical spectra web server}.} for isolated brown dwarfs in 
Fig.~\ref{fig:spectrum_comparison}. The "Helios-r2" results correspond to the modelling described earlier
in Sec.~\ref{sec:discussion.atmospheric_modelling} and the "VE21a" and "VE21b" results correspond to the modelling detailed in
\citet{VonEssen2021}. The rest of the results correspond to best-fitting BT-SETTL, BT-NextGen, BT-COND, BT-Dusty, and
\texttt{DRIFT-PHOENIX} models where the host star was modelled using a \texttt{BT-SETTL} spectrum with 
$T_\mathrm{Eff} = 6500$~K, $\log g = 4.0$ and $z = 0.0$. 

The best fitting BT and \texttt{DRIFT-PHOENIX} models correspond to a dayside temperatures of 3000~K (\texttt{DRIFT-PHOENIX}) 
and 3100~K (BT models) and fit even the \cheops passband surprisingly well. However, none of the isolated brown dwarf models can 
simultaneously explain both the \cheops and \tess eclipse depths.

\subsection{Sensitivity on parametrisation and priors}
\label{sec:discussion.sensitivity}

The choice of parametrisation and priors generally affects the parameter posteriors, and this effect can be
significant when studying weak signals. We parameterise the day- and nightside emission using log 
flux ratios and set uniform priors on these, which equals to setting log-uniform (reciprocal) priors on the day- and 
nightside emission flux ratios. This differs from previous studies using uniform priors on the flux ratios.

We repeated our final joint analysis parametrising it with flux ratios rather than log flux ratios to
study how sensitive our posteriors are on the choice of parametrisation. The most notable difference
was in nightside flux ratios, where uniform priors led to distributions with non-zero modes and the
maximum flux ratios (99th posterior percentile) were somewhat higher than when using log-uniform priors. 
The second notable difference was in the \cheops band dayside flux ratio posterior, which again 
had a non-zero mode with slightly larger maximum values. However, the difference in the maximum flux
ratio was not large enough to change the main outcome of the analysis. The dayside flux ratio posteriors 
for other passbands were not significantly affected.

We adopt the log flux ratio parametrisation since this should be more suitable for a "scale" parameter
with an unknown magnitude. This ensures that we do not give too much weight on apparent posterior modes
if in reality we can only estimate the upper limit securely.

\subsection{The role of ellipsoidal variation in the \cheops band}

Our \cheops band eclipse depth estimate is very sensitive on the amplitude of the ellipsoidal variation signal,
as illustrated in Fig.~\ref{fig:cheops_fr_scenarios} in Appendix~\ref{sec:analysis_scenarios}. However,
the two approaches leading to physically expected EV amplitudes yield similar \cheops band eclipse depths,
and only nonphysically low EV amplitudes can lead to eclipse depths that would agree with the \tess measurement.

\section{Conclusions}
\label{sec:conclusions}

We have estimated a self-consistent dayside emission spectrum of \nplanet covering the \cheops, \tess, \pbh, 
\pbks, and \spitzer IRAC 3.6 and 4.5~$\mu$m passbands by modelling new \cheops secondary eclipse photometry 
of \nplanet jointly with the existing ground- and space-based light curves. The study adds so-far the bluest
point to \nplanet's emission spectrum, improves the two ground-based NIR measurements by using a phase curve
model that includes a stellar ellipsoidal variation signal, and improves the \spitzer measurements by modelling
the \citet{Beatty2017} and \citet{Beatty2019} observations jointly with all the other light curves.

The emission spectrum largely agrees with the previous studies. The \pbh and \pbks dayside flux ratios are lower
than the prior estimates but this was expected since the previous modelling ignored the stellar ellipsoidal 
variation signal. The \tess and \spitzer 4.5~$\mu$m bands also agree with the results by \citet{Beatty2019}, 
\citet{Beatty2020}, and \citet{VonEssen2021}. The 3.6~$\mu$m \spitzer band dayside flux ratio estimate 
deviates both from previous results by \citet{Beatty2019} and theoretical expectations. This difference can
either be due to insufficiently modelled systematics or non-equilibrium chemistry effects in the brown dwarf
atmosphere, but we cannot securely determine its cause.

Another, significantly more difficult-to-explain discrepancy occurs between the \cheops and \tess passbands:
the \tess observations show a deep secondary eclipse with a depth of $360\pm60$~ppm, but the \cheops observations
strictly rule out an eclipse with a depth larger than 80~ppm (Fig.~\ref{fig:cheops_light_curve}).
The \cheops observations yield nearly twice higher photometric precision than the \tess observations, so
an eclipse similar to the \tess band one would be easily detected.

Atmospheric models can reproduce \nplanet's emission spectrum fairly well in most of the passbands considered,
but generally manage to manage to fit only either the \cheops or \tess band well.
Atmospheric modelling with \texttt{HELIOS} leads to a solution where \nplanet's geometric albedo is $\approx0.3$ in 
the \tess passband but consistent with 0 in the \cheops band. A contrast like this in the geometric albedo is difficult
to explain due to the overlap between the two passbands. 
The models by \citet{VonEssen2021} explain the \tess band without reflection but cannot explain the \cheops band 
eclipse depth (Fig.~\ref{fig:spectrum_comparison}). Finally, the BT and \texttt{DRIFT-PHOENIX} models \citep{Allard2013,Witte2011} 
for isolated brown dwarfs also reproduce the observed \nplanet emission spectrum fairly well, but have trouble
reconciling the \cheops and \tess eclipse depths.

It is challenging to explain the discrepancy between the \cheops and \tess band eclipse depths.
Temporal variability in cloud cover caused by the transport of transient nightside clouds to the dayside 
(as suggested by \citealt{Beatty2020}) could provide a potential explanation. High contrast features in 
\nplanet's albedo spectrum could also play a role in explaining a part of the discrepancy, and could
be caused by a layer of strong optical absorber such as gaseous TiO/VO or S$_2$/HS residing above 
a silicate cloud layer\citep{Schwartz2015}, but are unable to explain the discrepancy fully. 

All in all, instead of shedding light on which of the previously proposed theories might work best to explain
\nplanet's atmosphere, our additional blue-optical eclipse depth measurement has introduced a new open question
that is challenging to explain with our current knowledge. Further observations combined with comprehensive 
modelling are required to study whether \nplanet's dayside brightness spectrum is explained by transient silicate 
clouds \citep{Beatty2020}, thermal emission \citep{VonEssen2021}, or something else, and how the discrepancy 
between the \cheops and \tess band eclipse depths can be explained.

\begin{acknowledgements}
We thank the anonymous referee for their insight and useful comments.
We kindly thank T.G.~Beatty and B.~Croll for providing their observations to our use, and C.~von~Essen, M.~Mallonn, N.~Madhusudhan, and A.~Piette for providing their model spectra from \citet{VonEssen2021}.
CHEOPS is an ESA mission in partnership with Switzerland with important contributions to the payload and the ground segment from Austria, Belgium, France, Germany, Hungary, Italy, Portugal, Spain, Sweden, and the United Kingdom. The CHEOPS Consortium would like to gratefully acknowledge the support received by all the agencies, offices, universities, and industries involved. Their flexibility and willingness to explore new approaches were essential to the success of this mission. 
ACC and TGW acknowledge support from STFC consolidated grant numbers ST/R000824/1 and ST/V000861/1, and UKSA grant number ST/R003203/1. 
ML acknowledges support of the Swiss National Science Foundation under grant number PCEFP2\_194576. 
This project has received funding from the European Research Council (ERC) under the European Union’s Horizon 2020 research and innovation programme (project {\sc Four Aces}. 
grant agreement No 724427). It has also been carried out in the frame of the National Centre for Competence in Research PlanetS supported by the Swiss National Science Foundation (SNSF). DE acknowledges financial support from the Swiss National Science Foundation for project 200021\_200726. 
This work was supported by FCT - Fundação para a Ciência e a Tecnologia through national funds and by FEDER through COMPETE2020 - Programa Operacional Competitividade e Internacionalizacão by these grants: UID/FIS/04434/2019, UIDB/04434/2020, UIDP/04434/2020, PTDC/FIS-AST/32113/2017 \& POCI-01-0145-FEDER- 032113, PTDC/FIS-AST/28953/2017 \& POCI-01-0145-FEDER-028953, PTDC/FIS-AST/28987/2017 \& POCI-01-0145-FEDER-028987, O.D.S.D. is supported in the form of work contract (DL 57/2016/CP1364/CT0004) funded by national funds through FCT. 
LMS gratefully acknowledges financial support from the CRT foundation under Grant No. 2018.2323 ‘Gaseous or rocky? Unveiling the nature of small worlds’. 
S.G.S. acknowledge support from FCT through FCT contract nr. CEECIND/00826/2018 and POPH/FSE (EC). 
YA and MJH acknowledge the support of the Swiss National Fund under grant 200020\_172746. 
We acknowledge support from the Spanish Ministry of Science and Innovation and the European Regional Development Fund through grants ESP2016-80435-C2-1-R, ESP2016-80435-C2-2-R, PGC2018-098153-B-C33, PGC2018-098153-B-C31, ESP2017-87676-C5-1-R, MDM-2017-0737 Unidad de Excelencia Maria de Maeztu-Centro de Astrobiologí­a (INTA-CSIC), as well as the support of the Generalitat de Catalunya/CERCA programme. The MOC activities have been supported by the ESA contract No. 4000124370. 
S.C.C.B. acknowledges support from FCT through FCT contracts nr. IF/01312/2014/CP1215/CT0004. 
XB, SC, DG, MF and JL acknowledge their role as ESA-appointed CHEOPS science team members. 
ABr was supported by the SNSA. 
ACC acknowledges support from STFC consolidated grant numbers ST/R000824/1 and ST/V000861/1, and UKSA grant number ST/R003203/1. 
This project was supported by the CNES. 
The Belgian participation to CHEOPS has been supported by the Belgian Federal Science Policy Office (BELSPO) in the framework of the PRODEX Program, and by the University of Liège through an ARC grant for Concerted Research Actions financed by the Wallonia-Brussels Federation. 
L.D. is an F.R.S.-FNRS Postdoctoral Researcher. 
B.-O.D. acknowledges support from the Swiss National Science Foundation (PP00P2-190080). 
MF and CMP gratefully acknowledge the support of the Swedish National Space Agency (DNR 65/19, 174/18). 
DG gratefully acknowledges financial support from the CRT foundation under Grant No. 2018.2323 ``Gaseousor rocky? Unveiling the nature of small worlds''. 
M.G. is an F.R.S.-FNRS Senior Research Associate. 
SH gratefully acknowledges CNES funding through the grant 837319. 
KGI is the ESA CHEOPS Project Scientist and is responsible for the ESA CHEOPS Guest Observers Programme. She does not participate in, or contribute to, the definition of the Guaranteed Time Programme of the CHEOPS mission through which observations described in this paper have been taken, nor to any aspect of target selection for the programme. 
This work was granted access to the HPC resources of MesoPSL financed by the Region Ile de France and the project Equip@Meso (reference ANR-10-EQPX-29-01) of the programme Investissements d'Avenir supervised by the Agence Nationale pour la Recherche. 
PM acknowledges support from STFC research grant number ST/M001040/1. 
GSc, GPi, IPa, LBo, VNa and RRa acknowledge the funding support from Italian Space Agency (ASI) regulated by “Accordo ASI-INAF n. 2013-016-R.0 del 9 luglio 2013 e integrazione del 9 luglio 2015 CHEOPS Fasi A/B/C”. 
This work was also partially supported by a grant from the Simons Foundation (PI Queloz, grant number 327127). 
IRI acknowledges support from the Spanish Ministry of Science and Innovation and the European Regional Development Fund through grant PGC2018-098153-B- C33, as well as the support of the Generalitat de Catalunya/CERCA programme. 
GyMSz acknowledges the support of the Hungarian National Research, Development and Innovation Office (NKFIH) grant K-125015, a PRODEX Institute Agreement between the ELTE E\"otv\"os Lor\'and University and the European Space Agency (ESA-D/SCI-LE-2021-0025), the Lend\"ulet LP2018-7/2021 grant of the Hungarian Academy of Science and the support of the city of Szombathely. 
V.V.G. is an F.R.S-FNRS Research Associate. 
NAW acknowledges UKSA grant ST/R004838/1. 
M.F. gratefully acknowledge the support of the Swedish National Space Agency (SNSA; DNR 177/19, 174/18, and 65/19).
S.S. has received funding from the European Research Council (ERC) under the European Union’s Horizon 2020 research and innovation programme (grant agreement No 833925, project STAREX).
\end{acknowledgements}

\bibliographystyle{aa} 
\bibliography{kelt1b}

\appendix

\section{Chromatic albedo variability}
\label{sec:app.chromatic_albedo}
\subsection{Albedo spectrum model}

\begin{figure}
	\centering
	\includegraphics[width=\columnwidth]{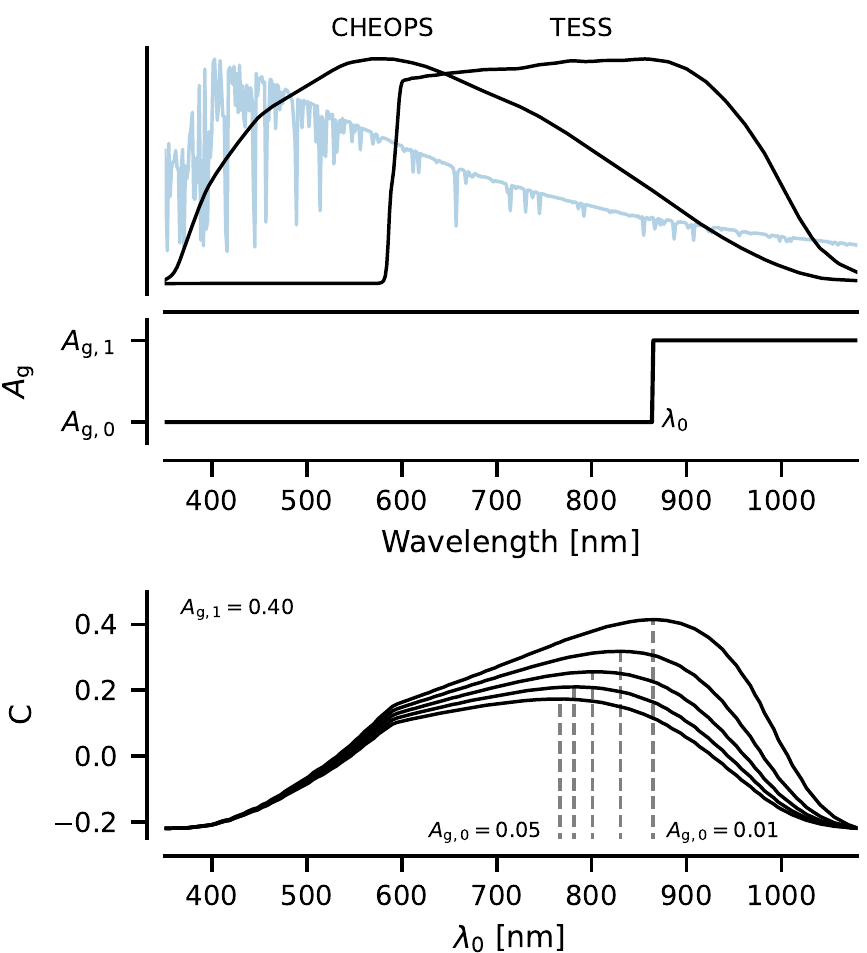}
	\caption{Top: a model spectrum of \nstar (light blue line), the \cheops and \tess passband transmission functions (black lines). Middle:
	the step-function albedo spectrum model. Bottom: \tess to \cheops reflected light contrast as a function 
    of \agb and $\lambda_0$. The step locations corresponding to the maximum flux ratios are marked as dashed vertical lines.}
	\label{fig:c_vs_t_albedo_1}
\end{figure}

We use a simple two-level albedo spectrum model to study whether the discrepancy between the \cheops and \tess band dayside flux
ratios could be caused by chromatic variability in albedo. The model is parametrised by two geometric albedo levels, \agb (blue)
and \agr (red), and a step location $\lambda_0$, so that
\begin{equation}
    \ag(\lambda) = 
    \begin{cases}
        \agb & \quad \text{if } \lambda < \lambda_0 \\
        \agr & \quad \text{if } \lambda \geq \lambda_0
    \end{cases},
\end{equation}
as illustrated in the middle panel of Fig.~\ref{fig:c_vs_t_albedo_1}. 

Our main quantity of interest is the contrast between the reflected light in the \cheops and \tess bands, 
\begin{equation}
    C= \frac{F_\mathrm{T} - F_\mathrm{C}}{F_\mathrm{T} + F_\mathrm{C}},\quad \text{where } F = \int S(\lambda)\, A_\mathrm{g}(\lambda)\, T(\lambda)\,\ud\lambda,
\end{equation}
$S$ is the stellar spectrum (a BT-Settl spectrum by \citealt{Allard2013}), \ag is the albedo spectrum model, 
and $T$ is the transmission function for either \tess ($F_\mathrm{T}$) or \cheops ($F_\mathrm{C}$).

We assume \agr that matches the geometric albedo for a "roaster" planet with silicate clouds from \citet{Sudarsky2000a},
compute $C$ for a set of \agb values between 0.01 (similar to TrES-2b, \citealt{Kipping2011b}) and 0.05, and $\lambda_0$ values between 400 and 1000~nmn
and show the results in the lower panel of Fig.~\ref{fig:c_vs_t_albedo_1}. The maximum $C$ is achieved for $\lambda_0 \sim 
800$~nm and varies from 0.17 to 0.41. We obtain $C=0.41$ for the lowest included blue \ag (as expected), $\agb=0.01$, and 
$C$ decreases quickly with increasing \agb.

\subsection{Comparison between modelled and observed reflection ratio}
\label{sec:app.chromatic_albedo.comparison}

Before we can compare our theoretical $C$ values to the $C$ distribution from the light curve analysis, we need to
estimate and remove the contribution from thermal emission in the latter. This contribution should be small for the \cheops 
and \tess bands, but is not necessarily negligible.

\begin{figure}
	\centering
	\includegraphics[width=\columnwidth]{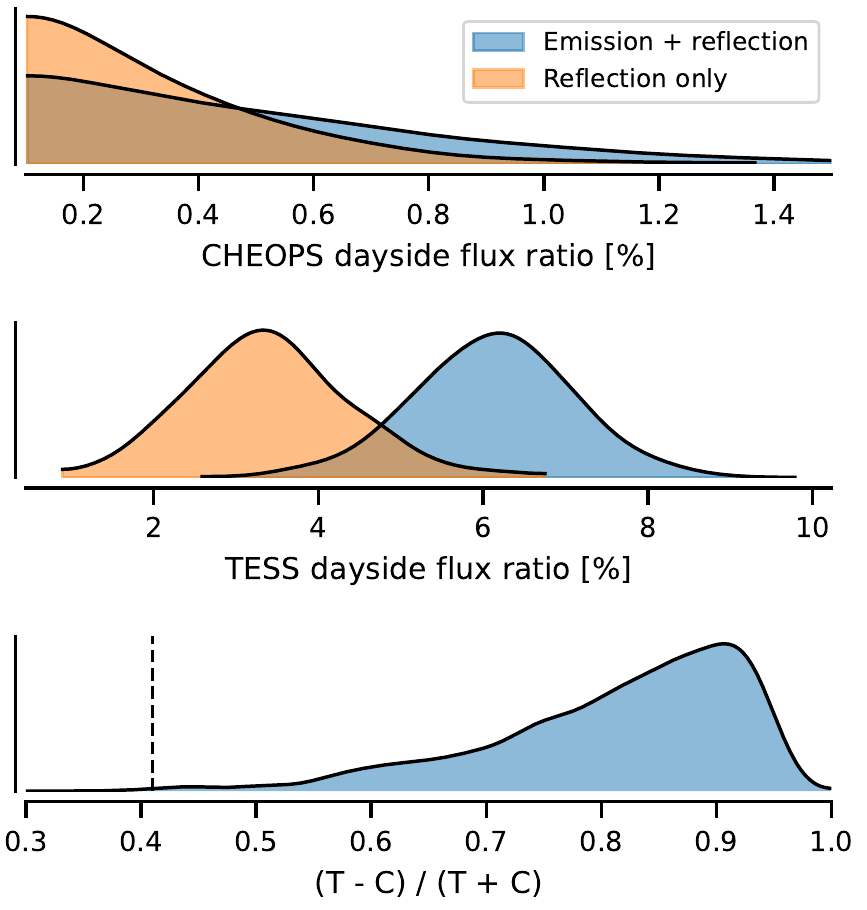}
	\caption{Top and middle: original \cheops and \tess dayside flux ratio distribution estimates together with versions of the 
                distributions with the thermal emission removed as described in Appendix~\ref{sec:app.chromatic_albedo.comparison}. 
                Bottom: the final \tess to \cheops contrast ($C$) distribution and the maximum $C$ of $\approx0.4$ from
                the albedo spectrum modelling marked with a dashed vertical line.}
	\label{fig:c_vs_t_albedo_2}
\end{figure}

We calculate a set of "reflection only" flux ratio samples from the original dayside flux ratio samples using a Monte Carlo approach.
We draw a set of \nplanet temperatures from a uniform distribution from 2800 to 3200~K (corresponding to the \nplanet \spitzer \pbsb
\tbr ranges given in Table~\ref{table:flux_ratio_posteriors}), we calculate the emission contribution in 
the \cheops and \tess bands using BT-Settl model spectra by \citet{Allard2013} for each temperature sample, and finally we remove 
these contributions from a random sample of \cheops and \tess dayside flux ratios . The full process is carried out in the 
\ghlink{A3_chromatic_albedo_variation_test.ipynb}{chromatic albedo variation test notebook}.
Figure~\ref{fig:c_vs_t_albedo_2} shows the original dayside flux ratios with emission and reflection, and the "reflection-only" flux
ratios. The correction is rather small for the \cheops band, but significant for the \tess band. The bottom panel of Fig.~\ref{fig:c_vs_t_albedo_2}
shows the final observed $R$ distribution. 

Our maximum $C$ of 0.41 is at the lower tail of the observed $C$ distribution. The maximum $C$ corresponds to a
low but still physically realistic red \ag of 0.01, which is similar to what has been measured for TrES-2b \citep{Kipping2011b}. However,
while both \agb and \agr are physically plausible separately, it is unclear if this is the case for an albedo spectrum combining 
the two.

\section{Comparison between EV scenarios}
\label{sec:analysis_scenarios}

In addition to the final joint analysis, we carried out a set of analyses for the external dataset (ED) consisting of the 
\tess, \pbh, \pbks, and \spitzer photometry, and another set of analyses for the \cheops photometry alone. The ED analysis was 
carried out to provide priors on the orbital and geometric parameters for the \cheops analysis and to study how different 
approaches to constrain the EV amplitudes affect our parameter estimates. The \cheops analysis was carried out to also study 
how the EV amplitude constraints affect the eclipse depth estimate, and to provide detrended \cheops light 
curves for the final joint analysis. 

Both the ED and \cheops-only analyses consider three scenarios that differ in the priors set on the EV amplitude: 
\begin{itemize}
    \item[a)] \emph{Theoretical EV:} the EV amplitudes are given normal priors based on theoretical estimates in 
    Table~\ref{table:ev_and_db_priors}. This is the most constraining scenario, but can lead to biased eclipse depth
    estimates if the semi-major axis estimate used to calculate the amplitudes is biased. 
    \item[b)] \emph{Constrained EV ratios:} the EV amplitudes are constrained relative to the EV amplitude in the
    \tess passband and the \tess passband EV amplitude is given a uninformative prior. This scenario works as a stepping 
    stone between the strongly constrained and completely unconstrained EV cases, and is not sensitive on our prior
    semi-major axis estimate.
    \item[c)] \emph{Unconstrained EV :} all the EV amplitudes are given uninformative priors to see whether the EV 
    amplitudes from the \tess and \spitzer passbands (where we have coverage over the full orbital phase) agree with 
    the theoretical expectations.
\end{itemize}
The \cheops-only analysis considers also an additional scenario with EV amplitude forced to zero:
\begin{itemize}
    \item[d)] \emph{No EV:} the EV amplitude is forced to zero. The motivation for this scenario is to see how ignoring 
    EV would affect our \cheops eclipse depth estimate.
\end{itemize}

Except for the data included, the models are identical to the full joint model. The baseline and noise models 
for the ED analyses are the same as in the final analysis. However, the baseline in the \cheops-only analysis is modelled 
differently than in the final analysis. In the \cheops analysis, we use the basis vectors determined by the pipeline 
by Wilson et al. as covariates and each visit is given its own set of baseline coefficients. The number of basis vectors 
per visit varies from 13 to 29, and the total number of basis vectors (and thus free parameters from the baseline model) 
is 156. This approach allows us to detrend the \cheops light curves together with the phase curve model to be used in the 
final joint analysis (this is safe and does not affect the final parameter estimates). The \cheops analysis uses the 
posteriors from the external dataset analysis as priors for the zero epoch, orbital period, stellar density, impact parameter, 
and planet-star area ratio. 

\begin{figure}
	\centering
	\includegraphics[width=\columnwidth]{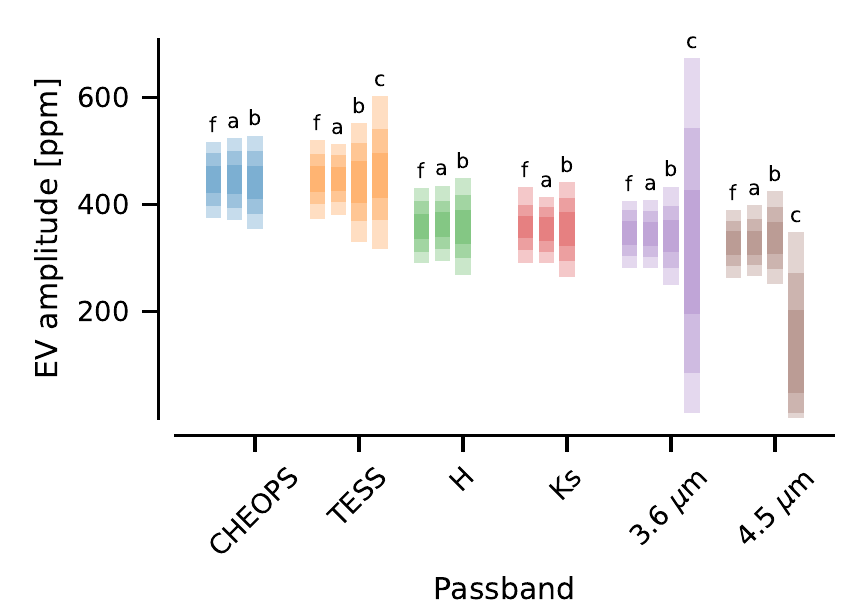}
	\caption{Ellipsoidal variation amplitude posteriors for f) final joint analysis with relative prior on the EV amplitude, 
	a) theoretical prior on EV amplitude, b) relative prior on EV amplitude, and c) uninformative prior on EV amplitude. 
	The case c with an uninformative prior on EV is shown only for the \tess and \spitzer passbands with full phase coverage.}
	\label{fig:ev_scenario_comparison}
\end{figure}

\begin{figure}
	\centering
	\includegraphics[width=\columnwidth]{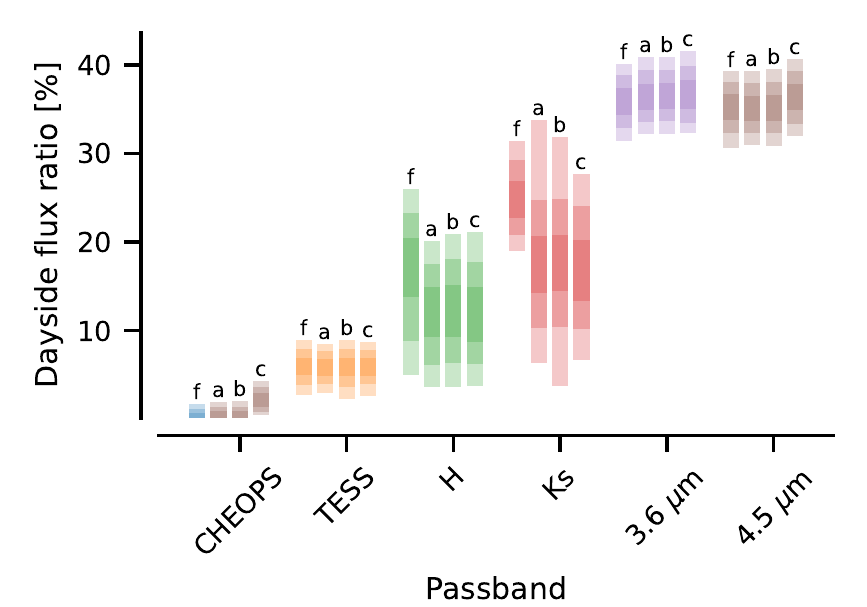}
	\caption{As in Fig.~\ref{fig:ev_scenario_comparison}, but for the dayside flux ratio.}
	\label{fig:all_fr_scenarios}
\end{figure}

\begin{figure}
	\centering
	\includegraphics[width=\columnwidth]{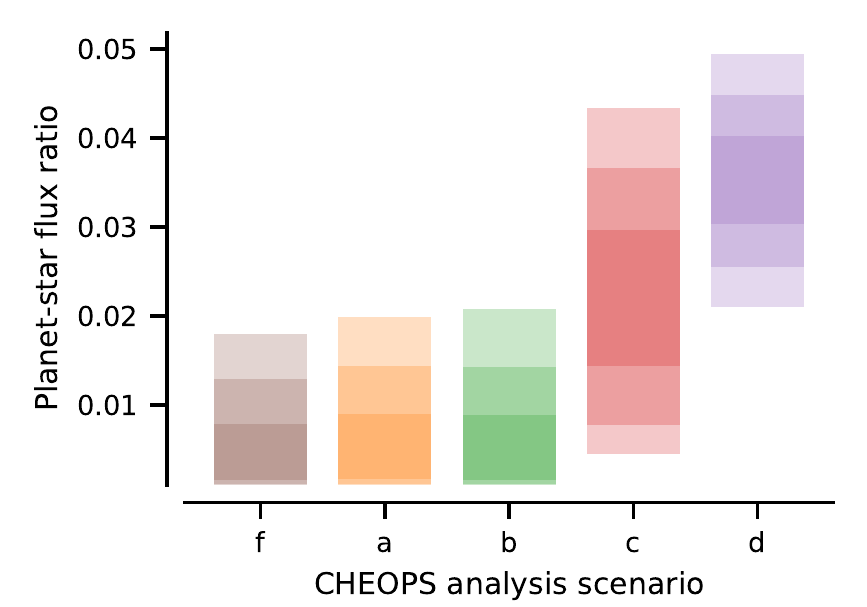}
	\caption{Dayside flux ratios the \cheops-only analysis scenarios. The scenario labels are the same as in 
	Fig.~\ref{fig:ev_scenario_comparison} with an additional no-EV scenario d.}
	\label{fig:cheops_fr_scenarios}
\end{figure}

We show the EV amplitude posteriors for all the analyses in Fig.~\ref{fig:ev_scenario_comparison}, the dayside flux ratio 
posteriors in Fig.~\ref{fig:all_fr_scenarios}, and the \cheops dayside posteriors in Fig.~\ref{fig:cheops_fr_scenarios}.
All in all, the dayside flux ratio posteriors all agree with each other for the final joint modelling and EV prior scenarios
a and b. The flux ratio posteriors also agree for scenario d for passbands with photometry covering a full orbital phase,
but the free- and no-EV scenarios (c and d) lead to overestimated flux ratios for \cheops photometry.

\section{Comparison between brightness temperature calculation approaches}
\label{sec:app.brightness_temperatures}

We calculate the brightness temperatures discussed in Sect.~\ref{sec:discussion.brightness_temperatures} using three 
approaches: physical-physical, where both the star and the brown dwarf are modelled using the BT-Settl spectra by 
\citet{Allard2013}; blackbody-physical, where \nplanet is modelled as a black body and its host star using a BT-Settl 
spectrum; and blackbody-blackbody, where both the brown dwarf and its host star are modelled as black bodies. Our 
reported brightness temperature values correspond to averages over the three approaches, but we also tested whether
the different approaches agree with each other or not.

\begin{figure}
	\centering
	\includegraphics[width=\columnwidth]{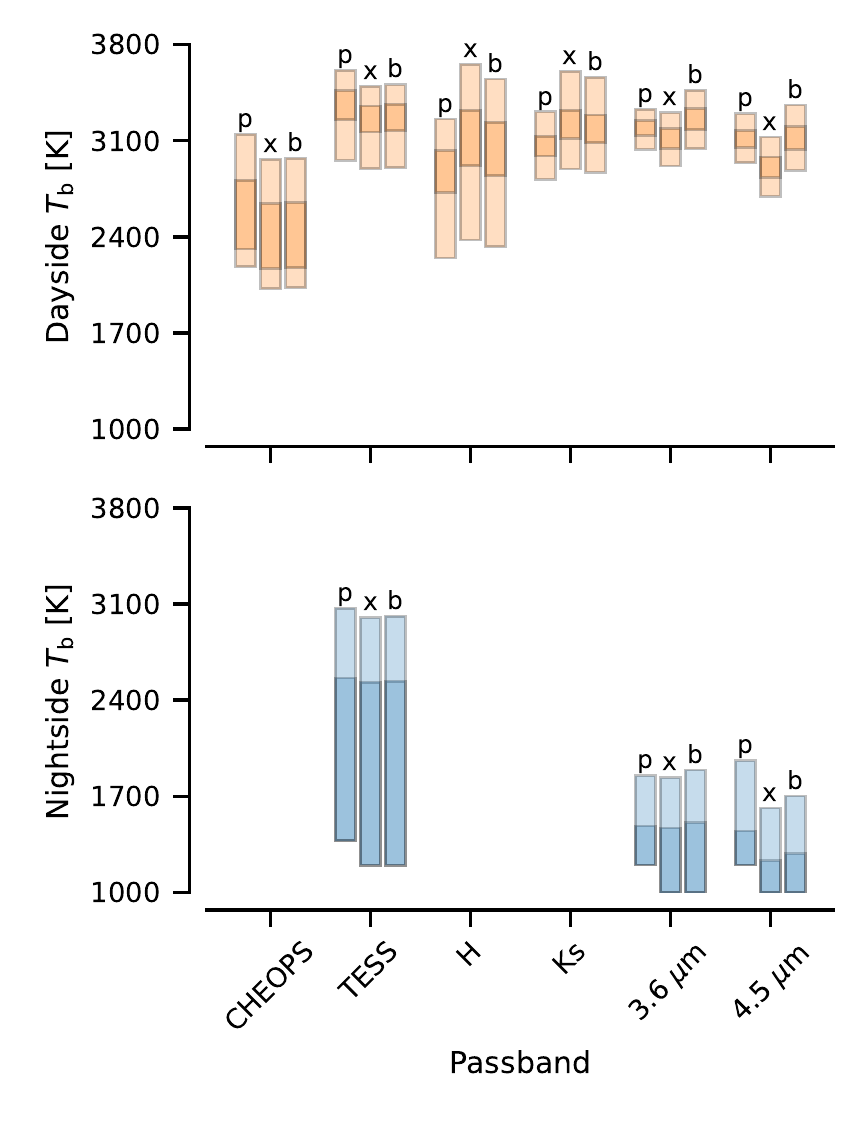}
	\caption{Day- and nightside brightness temperatures estimated from the flux ratios using the p) physical-physical,
	x) blackbody-physical, and b) blackbody-blackbody approaches. The shading marks the 68\% and 99.7\% central posterior 
	intervals.}
	\label{fig:brigthness_temperatures_sep}
\end{figure}

We present the day- and nightside brightness temperatures separately for each approach in Fig.~\ref{fig:brigthness_temperatures_sep}.
All the three approaches agree with each other within uncertainties.

\section{Additional retrievals to explore dependence of outcomes on data}
\label{sec:additional_retrievals}

\begin{figure*}
	\centering
	\includegraphics[width=0.48\textwidth]{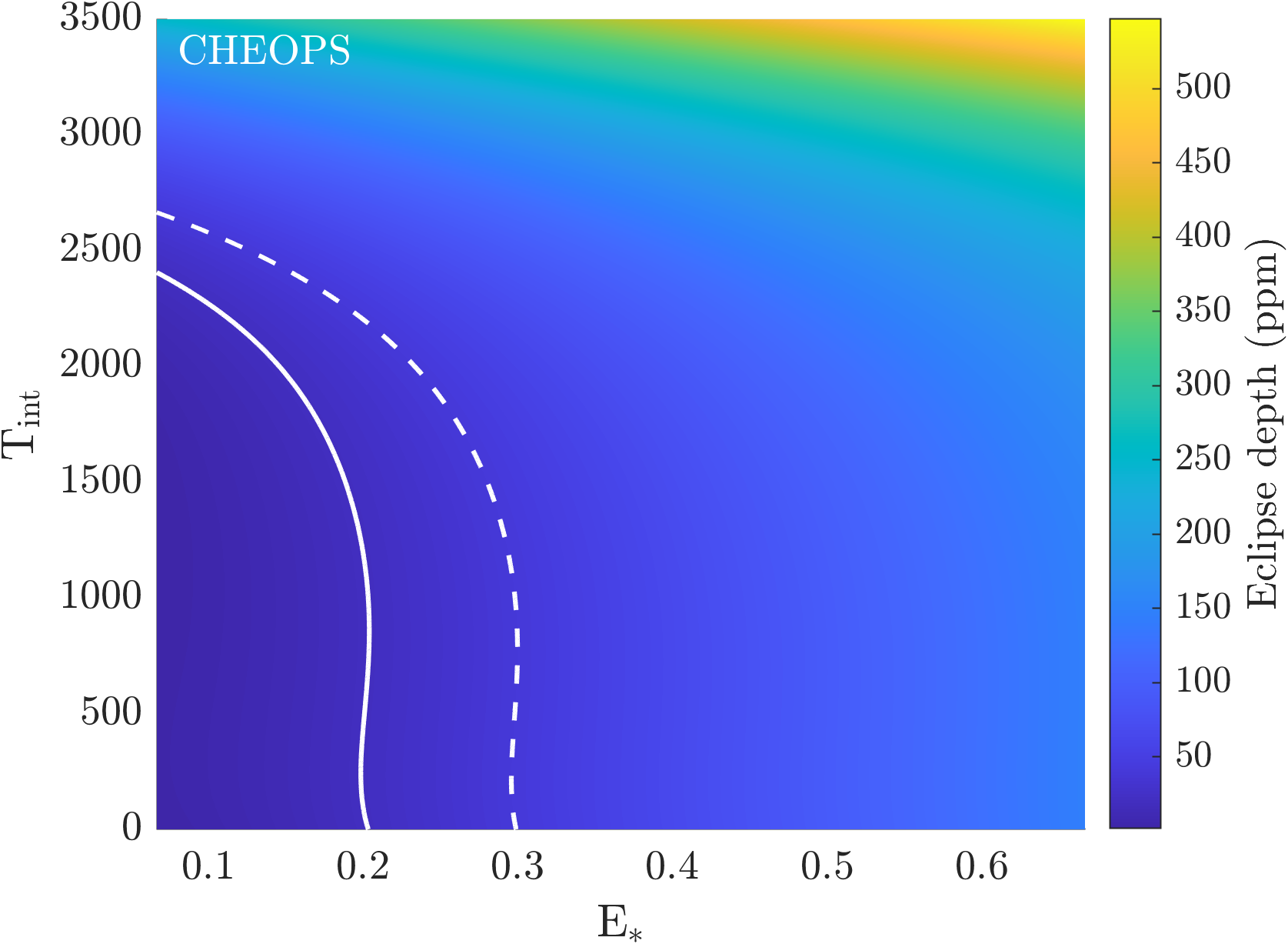}
	\includegraphics[width=0.48\textwidth]{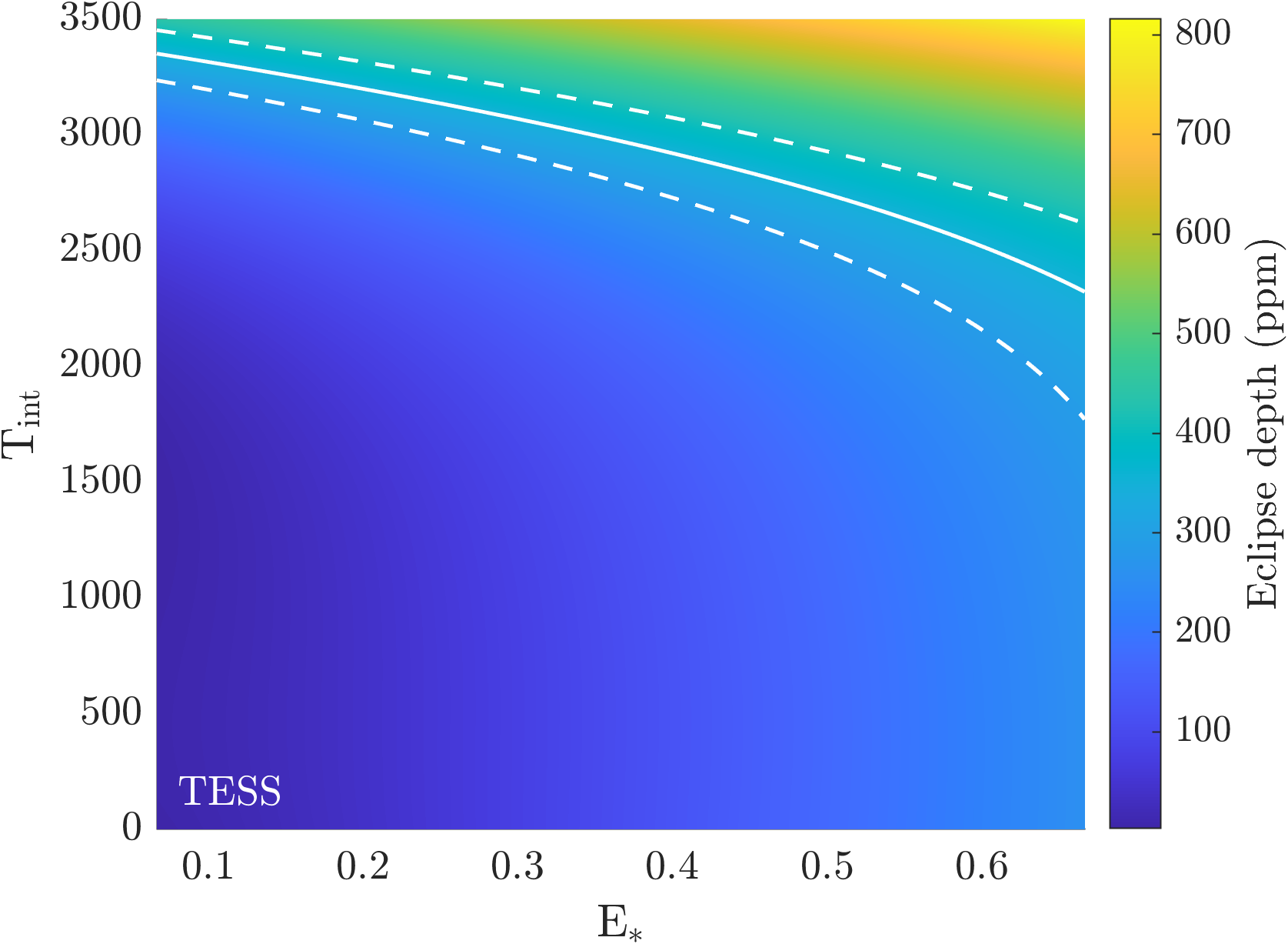}\\
	\includegraphics[width=0.48\textwidth]{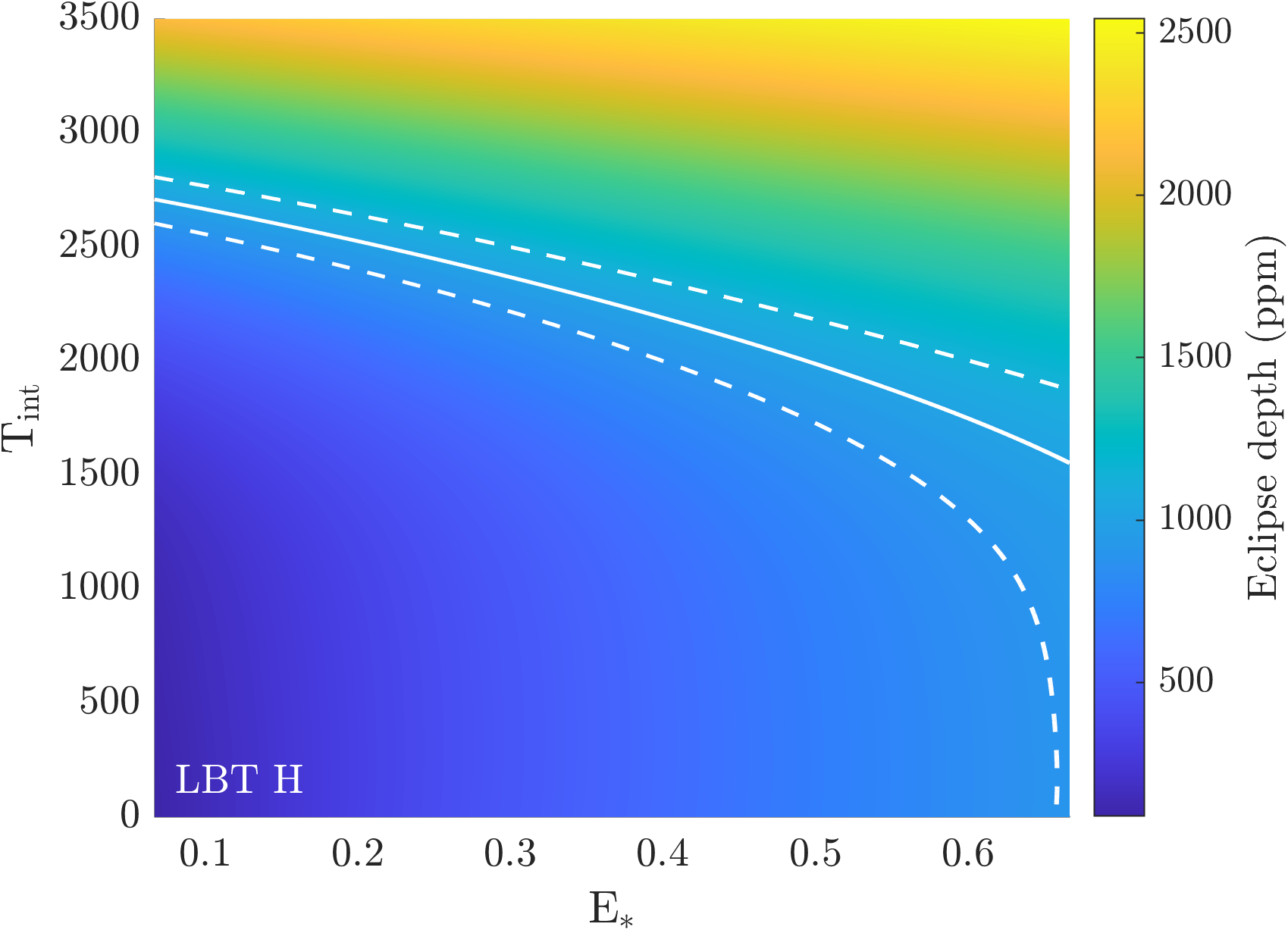}
	\includegraphics[width=0.48\textwidth]{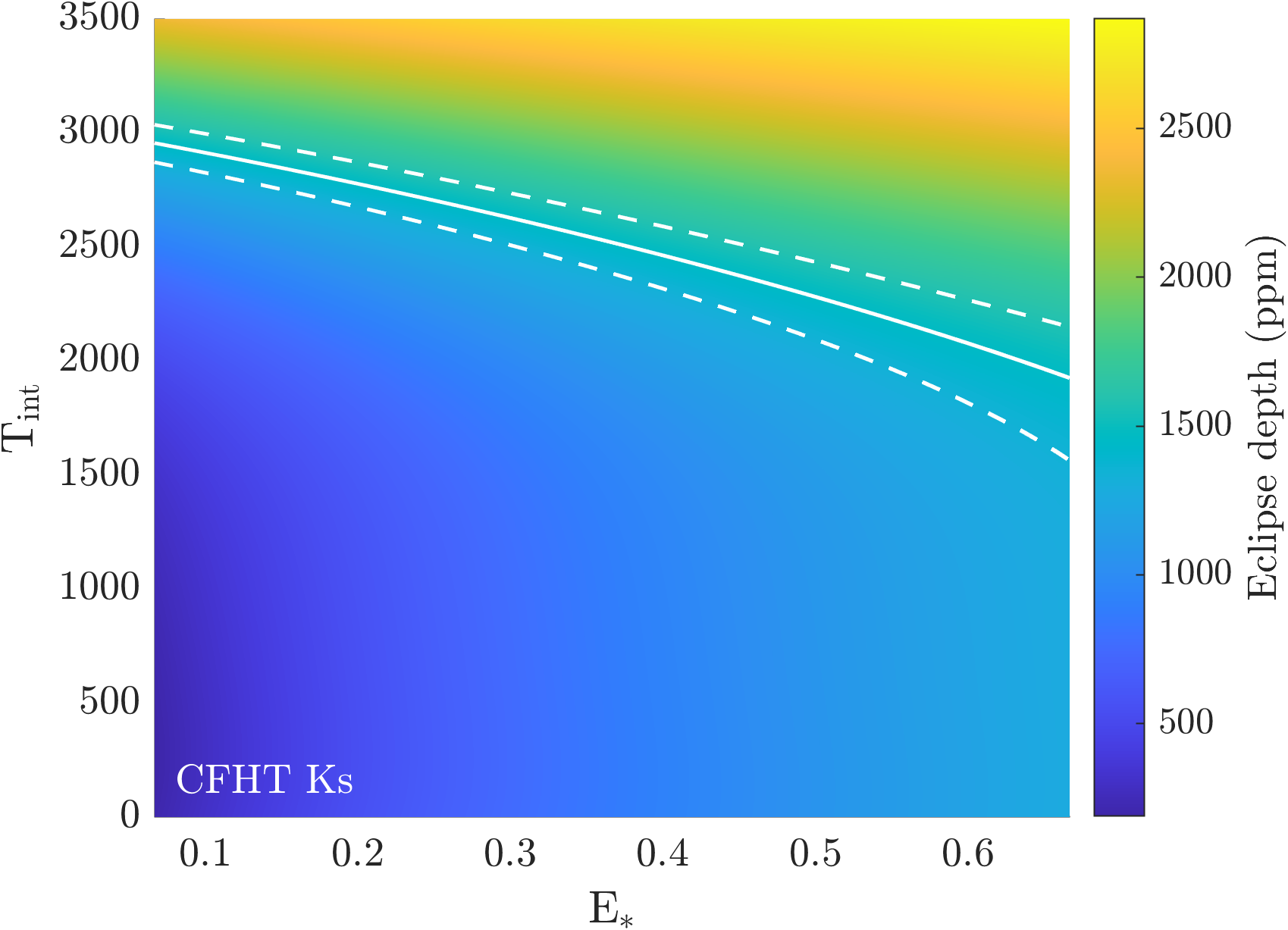}\\
	\includegraphics[width=0.48\textwidth]{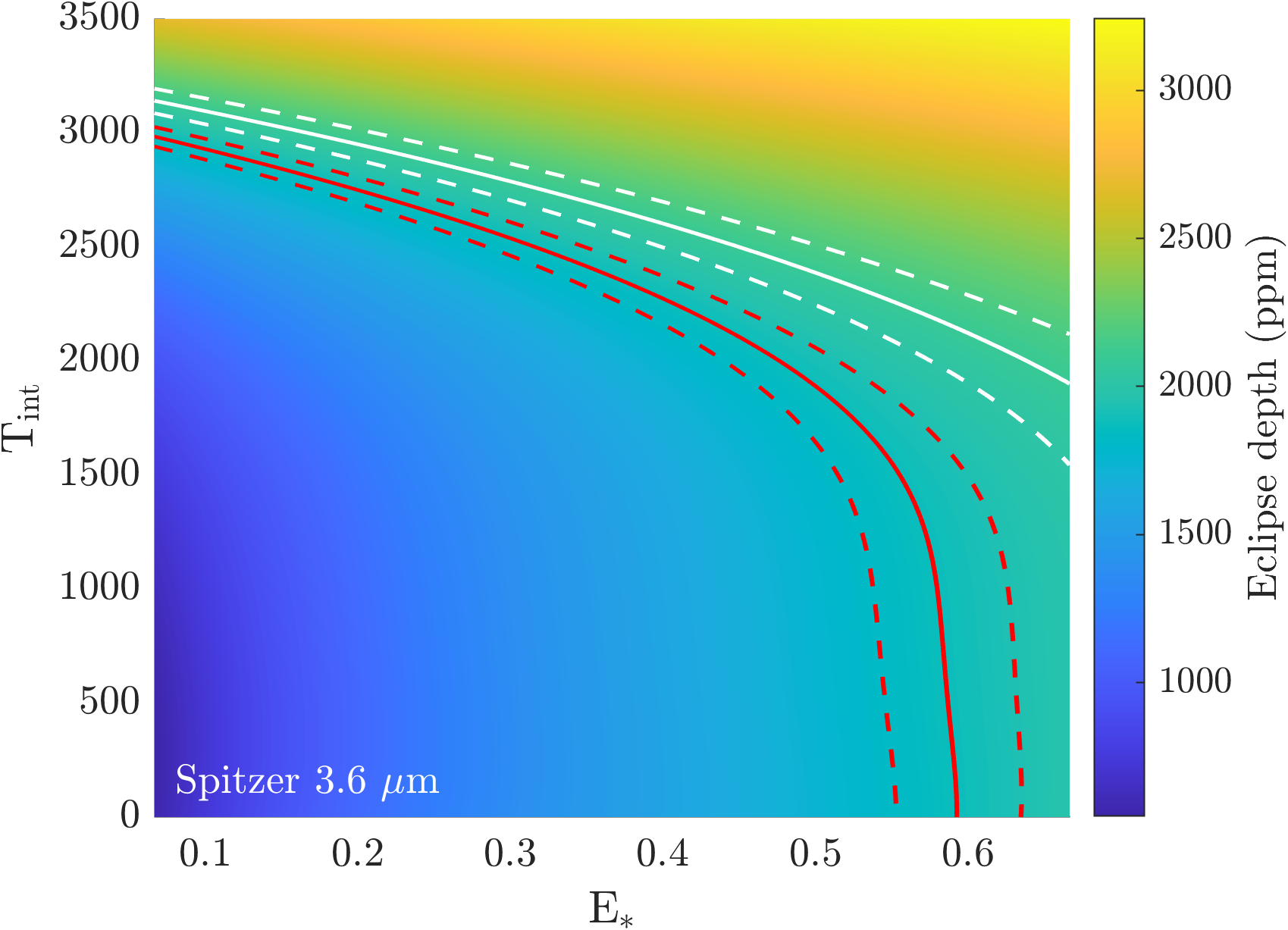}
	\includegraphics[width=0.48\textwidth]{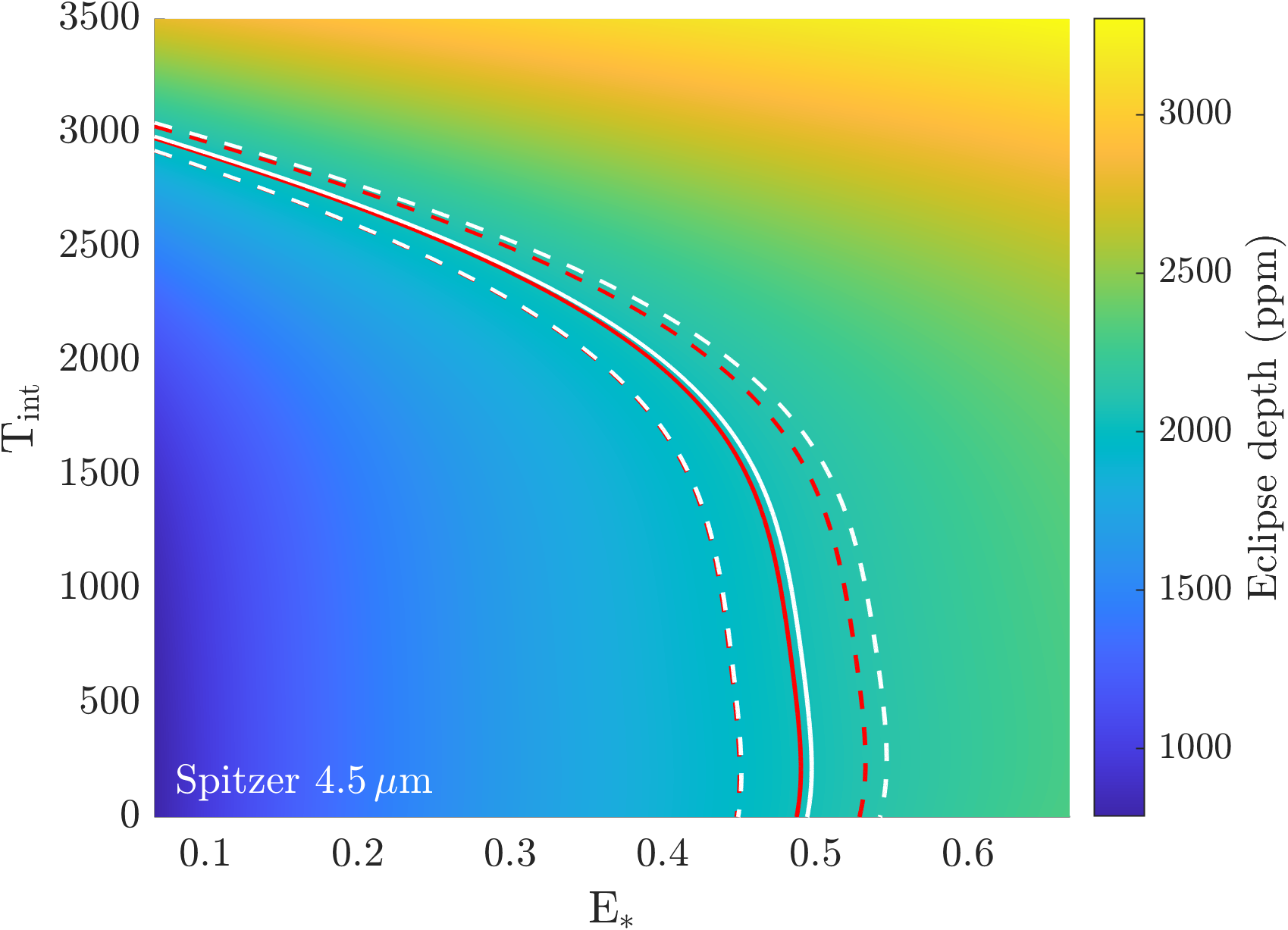}
	
	\caption{Eclipse depths in ppm as a function of the internal temperature $T_\mathrm{int}$ and the parameter $E_*$ based on the \texttt{HELIOS} model calculations described in Sect. \ref{sec:discussion.atmospheric_modelling}. The six panels show the eclipse depths in the various passbands. White, solid lines depict the measured occultation depths from Table \ref{table:flux_ratio_posteriors}, dashed lines refer to their $1\sigma$ error bars. The red lines for the two Spitzer passbands refer to the eclipse depths reported by \citet{Beatty2019}.}
	\label{fig:helios_secondary_eclipse}
\end{figure*}

Figure \ref{fig:helios_secondary_eclipse} shows the degeneracy between $T_{\rm int}$ and $E_*$ when the model accounts for the \tess secondary eclipse.  Within this family of solutions, the solution of \cite{VonEssen2021}, which explains the dayside emission spectrum without the need for reflected light, is included.

\begin{figure*}
	\centering
	\includegraphics[width=0.32\textwidth]{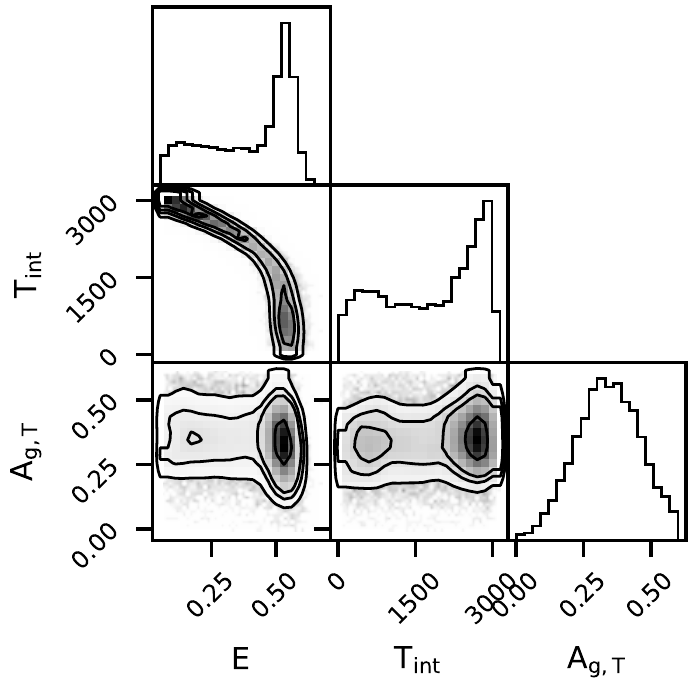}
	\includegraphics[width=0.32\textwidth]{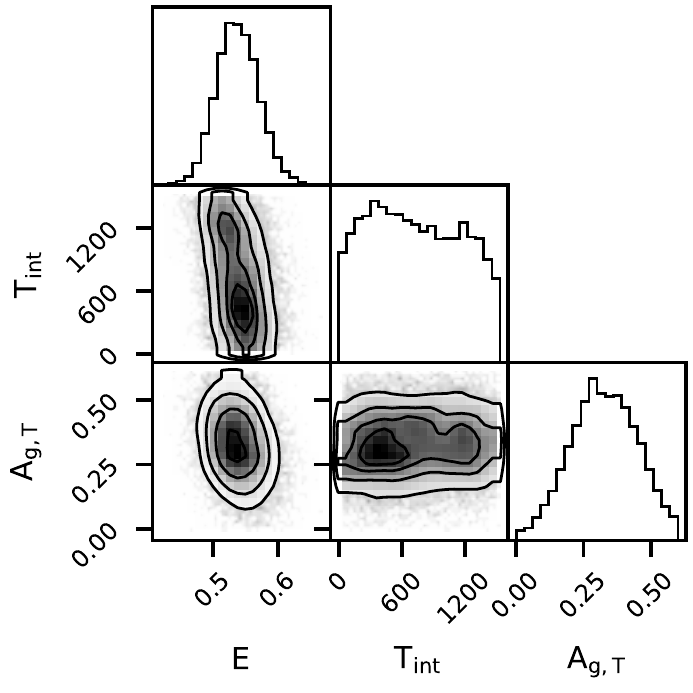}
	\includegraphics[width=0.32\textwidth]{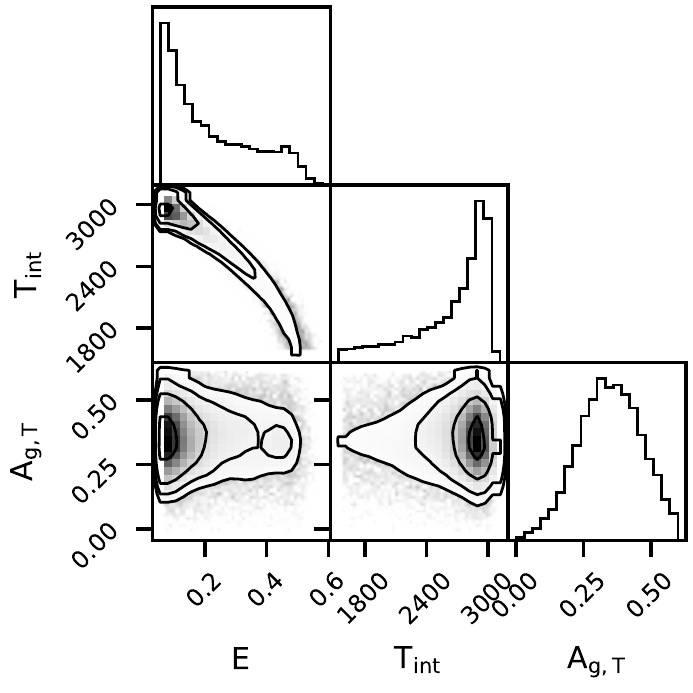}
	\caption{Posterior distributions for a retrieval with \texttt{Helios-r2} using only the \tess secondary eclipse measurement and the two \spitzer points from \citet{Beatty2019}. The left panel shows the results for the full range of prior distributions discussed in Sect. \ref{sec:discussion.atmospheric_modelling}, which yields a bimodal solution with respect to $T_\mathrm{int}$ and the parameter $E_*$. The two other panels isolate the two solution modes.}
	\label{fig:posterior_tess_spitzer_beatty}
\end{figure*}

In Fig. \ref{fig:posterior_tess_spitzer_beatty}, we perform a retrieval with just the \tess, \spitzer 3.6 $\mu$m and 4.5 $\mu$m data points reported by \cite{Beatty2019}.  The left panel of \ref{fig:posterior_tess_spitzer_beatty} shows a a retrieval with the full prior ranges. This yields a bimodal solution for $T_{\rm int}$ and $E_*$: a high value of $E_*$ and a low value of $T_{\rm int}$, or a low value of $E_*$ and a high value of $T_{\rm int}$.  The two different solutions are isolated in the other two panels of Fig. \ref{fig:posterior_tess_spitzer_beatty}.

When the \pbh and \pbks data points are added to the retrieval, the bimodality is broken (not shown). This can be understood by looking at the ranges of $T_{\rm int}$ and $E_*$ values that are consistent with the measurements in this passbands depicted in  \ref{fig:helios_secondary_eclipse}. The results presented in this figure suggest that the \pbh and \pbks secondary eclipse measurements are only consistent with high values of $T_{\rm int}$ 

\begin{figure}
	\centering
	\includegraphics[width=\columnwidth]{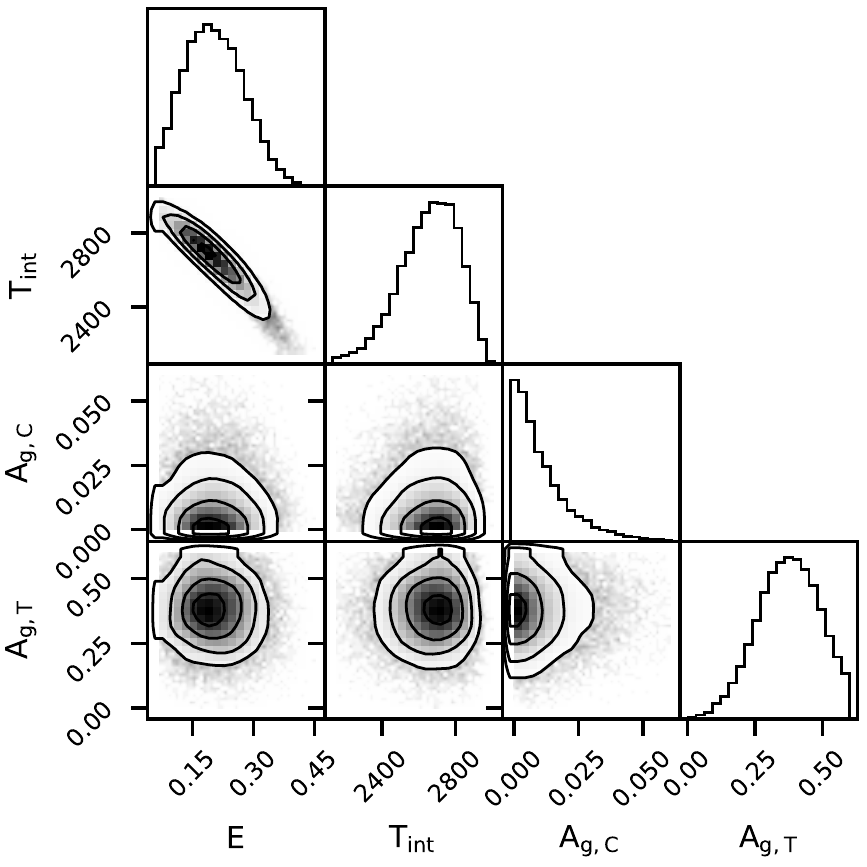}
	\caption{Posterior distributions for a retrieval with \texttt{Helios-r2} using all secondary eclipse measurement from Table \ref{table:flux_ratio_posteriors} but with the two \spitzer points from \citet{Beatty2019}.}
	\label{fig:posterior_all_beatty}
\end{figure}

Finally, in Fig. \ref{fig:posterior_all_beatty} we perform a retrieval using all available measurements but with the \spitzer secondary eclipse depths from \cite{Beatty2019}. The addition of the \cheops data point now further narrows the posterior on $E_*$ down to a median value of 0.2. this behaviour can also be easily understood by comparison with the theoretically calculated eclipse depths in Fig. \ref{fig:helios_secondary_eclipse}. For \cheops, the results indicate that the measured eclipse depth is only compatible with a maximum $E_*$ of about 0.2. 

We also note that the results using the \spitzer data from \cite{Beatty2019} essentially provides the same posterior distribution than using the reduction of the \spitzer of this work (see Fig. \ref{fig:retrieval_posteriors}). . In particular, the very high geometric albedo in the \tess passband is consistently obtained with either dataset. Despite the difference in the reported \spitzer 3.6 $\mu$m eclipse depth, the retrieval outcome is, thus, unaffected because the solution with $T_{\rm int} \approx 2700$~K has a blackbody peak at shorter wavelengths, within the \pbh and \pbks passbands.

\end{document}